\documentclass[prd, twocolumn, amsmath, floatfix, preprintnumbers, nofootinbib,superscriptaddress]{revtex4-1}

\usepackage{graphicx}
\usepackage{amsbsy}
\usepackage{amssymb}
\usepackage{dsfont}
\usepackage{layouts}
\usepackage{mdframed}
\usepackage{natbib}
\usepackage{xcolor}
\usepackage{mathtools}
\usepackage{subfigure}
\usepackage{dcolumn}
\usepackage{units}

\newcolumntype{.}{D{.}{.}{-1}}

\usepackage[colorlinks,citecolor=red,linkcolor=blue]{hyperref}

\def\cqg{Class. Quantum Grav.}
\def\aap{Astron. Astrophys.}

\newcommand{\X}{\mathbf{X}}
\newcommand{\Y}{\mathbf{Y}}
\newcommand{\Zt}{\mathbf{Z}^{\dagger}}
\newcommand{\B}{\mathbf{B}}
\newcommand{\Best}{\mathbf{\hat{B}}}

\newcommand{\R}{\mathbf{R}}
\newcommand{\M}{\mathbf{M}}
\newcommand{\Sig}{\mathbf{\Sigma}}
\newcommand{\Cx}{\mathbf{C}}

\newcommand{\yv}{\mathbf{y}}
\newcommand{\xv}{\mathbf{x}}
\newcommand{\sv}{\mathbf{s}}
\newcommand{\bv}{\mathbf{b}}
\newcommand{\mv}{\mathbf{m}}
\newcommand{\ev}{\mathbf{e}}

\begin{document}
  \title{Multivariate Regression Analysis of Gravitational Waves from Rotating
Core Collapse}

  \preprint{\texttt{LIGO-P1300216}}
  \date{\today}
  \author{William J. Engels}
  \email{wengels@uoregon.edu}
  \affiliation{Department of Physics, University of Oregon, Eugene, OR, USA}
  \author{Raymond Frey}
  \email{rayfrey@uoregon.edu}
  \affiliation{Department of Physics, University of Oregon, Eugene, OR, USA}
  \author{Christian D. Ott}
  \affiliation{TAPIR, California Institute of Technology, Pasadena, CA, USA\\
  Kavli Institute for the Physics and Mathematics of the Universe (Kavli IPMU; WPI), Kashiwa, Japan}
  \email{cott@tapir.caltech.edu}

\begin{abstract}
We present a new multivariate regression model for analysis and
parameter estimation of gravitational waves observed from well but not
perfectly modeled sources such as core-collapse supernovae. Our
approach is based on a principal component decomposition of simulated
waveform catalogs. Instead of reconstructing waveforms by direct
linear combination of physically meaningless principal components, we
solve via least squares for the relationship that encodes the
connection between chosen physical parameters and the principal
component basis. Although our approach is linear, the waveforms'
parameter dependence may be non-linear. For the case of gravitational
waves from rotating core collapse, we show, using statistical hypothesis
testing, that our method is capable of identifying the most important
physical parameters that govern waveform morphology in the presence of
simulated detector noise. We also demonstrate our method's ability to
predict waveforms from a principal component basis given a set of
physical progenitor parameters.
\end{abstract}

\maketitle

\section{Introduction}
\label{section:intro}

Unimpeded by intervening material, gravitational waves (GWs) trace out
bulk motions of matter in the sudden collapse of a dying massive
star's core~\cite{Ott2009}.  Hidden beneath the stellar envelope,
these dynamics are inaccessible by traditional observational methods.
After the star's iron core exceeds its effective Chandrasekhar mass,
it grows gravitationally unstable and collapse ensues.  The stiffening
of the nuclear equation of state (EOS) at nuclear density leads to the
rebound of the inner core (``core bounce'') into the still infalling
outer core, creating an outwardly propagating shock wave.  According
to simulations and basic theory (e.g., \cite{bethe:90}), this shock
wave quickly deteriorates and is not sufficiently energetic enough to
expel the stellar material and drive a supernova explosion. Instead,
it stalls and turns into an accretion shock. The yet uncertain
\emph{supernova mechanism} must revive the stalled shock. All
currently discussed candidate mechanisms involve multi-dimensional
bulk motions of matter in the region behind the stalled shock (e.g.,
\cite{janka:12b}). Hence, the detection, analysis, and
characterization of gravitational waves (GW) from core-collapse
supernovae could potentially provide great insights into the uncertain
mechanism that reignites the explosion.

As supernova theorists converge on accurate models to describe and
predict the transition from core collapse to supernova explosion,
advanced GW detectors such as Advanced LIGO~\cite{LIGO} and Advanced
Virgo~\cite{VIRGO} will begin taking data with $\sim$ten times greater
sensitivity than their initial versions.  Since the expected rate of
galactic core-collapse supernovae is only $\sim 1-3$ per
century~(e.g., \cite{Adams2013}), it is imperative to develop methods
able to extract as much information as possible from the GWs that will
be observed from these rare events.

Theory and multi-dimensional simulations have identified a variety of
GW emission processes, including rotating core collapse,
nonaxisymmetric rotational instabilies, turbulent convection in the
protoneutron star and in the region immediately behind the stalled
shock, pulsations of the protoneutron star, and asymmetric outflows of
mass-energy (see, e.g., \cite{Ott2009,kotake:13review} for
reviews). Of these emission processes, rotating core collapse is the
most extensively studied and has received the most attention from GW data
analysts.

In previous work, Brady and Majumdar~\cite{Brady2004} introduced a
Gram-Schmidt method to parameterize rotating core collapse GW signals
in terms of small numbers of orthonormal basis vectors encapsulating
robust signal features extracted from a catalog of simulated waveforms
by \cite{zwerger:97}. Heng~\cite{Heng2009} applied Principal Component
Analysis (PCA) for the same purpose, and showed that the PC basis (PCs;
principal components)
provides a more efficient representation of waveform catalogs than
Gram-Schmidt.

Summerscales~\emph{et al.}~\cite{Summerscales} studied the
reconstruction of rotating core collapse waveforms of \cite{ott:04}
injected into detector noise using a maximum entropy approach.  They
used cross-correlation of the reconstructed signal with catalog
waveforms to determine parameters of the source.

R\"over~\emph{et al.}~\cite{Rover2009} combined the PC basis approach
of \cite{Heng2009} with Bayesian inference (via Markov Chain Monte
Carlo) to recover the linear combination of PC basis vectors that most accurately reconstructs a rotating
core collapse GW signal buried in noise. They then compared the
recovered linear combination coefficients to the coefficients
associated with the rest of the catalog signals to infer the physical
parameters of the detected signal in a nearest-neighbor-type scheme
\cite{Tibshirani}.  While able to produce excellent reconstructions,
they had limited success inferring the physical parameters of the
recovered waveform.

Different explosion mechanisms may have distinct and characteristic GW
signatures \cite{Ott2009,ott:09b}. Exploiting this possibility,
Logue~\emph{et al.}~\cite{Logue2012} developed a Bayesian model
selection framework with the aim of inferring the explosion mechanism
on the basis of a GW signal in the presence of detector noise. They
used PC-decomposed waveform catalogs from simulations addressing
various GW emission models and computed the Bayesian evidence to infer
which catalog best reconstructs an injected signal.

The above previous work has demonstrated that PCA is a powerful tool
to extract robust features from an ensemble of waveforms modeling
different realizations (random realizations and/or variations of model
parameters) of the same GW emission process. However, as already noted
by \cite{Heng2009,Rover2009,Logue2012}, PCA's major disadvantage is
that the PCs do not directly encode the \emph{physical parameters} of
the simulated collapse models whose GW waveforms they represent. This
is a major limitation to their application in Bayesian inference
beyond model selection.

In this paper, we present a multivariate regression approach that
expresses the set of waveforms in a given core-collapse supernova GW
catalog as a linear combination of vectors, each corresponding to
features \emph{directly} attributable to progenitor characteristics.
Each of these waveform feature vectors is subsequently expressed as a
linear combination of PCs, providing a bridge between physical
parameters and PCs that is missing in previous work.  This method of
decomposing a waveform catalog allows us to characterize linear and
non-linear relationships between waveforms and physical parameters.

A similar multivariate regression approach was first used by Potthoff and
Roy~\cite{Potthoff1964} to conduct an analysis of variance of growth
curves.  Instead of a PC basis, they used a polynomial basis to study
the influence of different treatments on the growth of animal subjects
over time.  Zerbe and Jones~\cite{Zerbe1980} used a Fourier basis to
analyze circadian rhythm data.  Using the rotating core collapse
waveform catalog of Abdikamalov~\emph{et al.}~\cite{Abdikamalov2013},
we show that the statistical significance of these relationships can
be assessed via standard test statistics. By operating in the Fourier
domain, we can straightforwardly take corrupting detector noise into
account in these tests.

While we concentrate on applying our approach in an analysis of the
relationships between physical parameters and waveform features for
rotating core collapse, we also demonstrate that the method presented
can be used to construct rotating core collapse gravitational waveform
predictions using physical parameters as input. This work thus paves
the way for a template-bank based parameter estimation approach
for gravitational waves from rotating core collapse.

This paper is structured as follows. In Sec.~\ref{section:methods}, we
introduce the motivating rotating core collapse waveform catalog and
develop a statistical model for its analysis.  In
Sec.~\ref{sec:AbCat}, we review the physical parameter space used in
the Abdikamalov~\emph{et al.} waveform catalog.
In~Secs.~\ref{sec:MethodsOverview}
and~\ref{section:constructingthemodel}, we detail the steps we take to
mathematically describe a linear relationship between the
gravitational waveforms, features associated with physical parameters
and additive detector noise.  Sections~\ref{section:encoding}
and~\ref{section:basis} elaborate on how physical parameters are
encoded into our statistical model and our use of the SVD basis to
construct feature vectors.  In Sec.~\ref{section:solutions}, we
provide least squares solutions which estimate the feature vectors and
their covariances.  In Secs.~\ref{sec:statistics} through
Sec.~\ref{sec:interactions}, we present an analysis of the
relationships between physical parameters and the waveforms of the
Abdikamalov~\emph{et al.} core-collapse waveform catalog.  Finally in
Sec.~\ref{sec:polyresults}, we use our multivariate model to construct
waveforms not previously included in the analysis, and then compare
our predictions to the actual waveforms simulated by
Abdikamalov~\emph{et al.} in Sec.~\ref{sec:oos}.

\section{Methods and Inputs}
\label{section:methods}

\subsection{The Abdikamalov~\emph{et al.} Waveform Catalog}
\label{sec:AbCat}

Rapid rotation, in combination with strong magnetic fields, has been
suggested to enable a \emph{magnetorotational mechanism} for
core-collapse supernova explosions (e.g.,
\cite{bisno:70,burrows:07b}). In this mechanism, angular momentum
conservation leads to a rapidly differentially spinning postbounce
core. The magnetorotational instability (MRI; e.g., \cite{balbus:91})
is invoked to extract differential rotation energy and produce a local
magnetar-strength magnetic field. Depending on the initial rotation
rate (which should be fast enough to make a millisecond-period
protoneutron star) and the presence of a dynamo process that converts
local unordered field into global field, toroidal field strength of up
to $10^{15} - 10^{16}\,\mathrm{G}$ may be obtained.  If this is indeed
the case, a number of axisymmetric (2D) simulations have shown that
strong bipolar jet-like outflows develop that drive an explosion
(e.g., \cite{bisno:70,burrows:07b,takiwaki:11}). Recent full 3D
simulations reported in \cite{moesta:14b} suggest that in 3D the jet
is distorted by nonaxisymmetric instabilities and if an outflow
develops, it will not be as neatly collimated as in the 2D case.

A rapidly rotating core has a natural quadrupole moment due to its
flattening by the centrifugal force. The extreme accelerations at core
bounce lead to a rapid and large-scale change in the quadrupole
moment. This gives rise to a characteristic GW signal that is
predominantly linearly polarized (e.g.,
\cite{ott:07prl,scheidegger:10}). This signal is so distinct from
other GW emission processes in core-collapse supernovae that it is
possible to use it as an indicator for the rapid rotation required for
magnetorotatoinal explosions \cite{Ott2009,ott:09b,Logue2012}.

Abdikamalov~\emph{et al.}~\cite{Abdikamalov2013} recently carried out
135 axisymmetric general-relativistic hydrodynamic simulations of
rotating core collapse\footnote{The Abdikamalov~\emph{et al.} waveform
  catalog is available at
  \url{http://stellarcollapse.org/ccdiffrot}.}. Since the GW signal
from rotating core collapse is essentially independent of progenitor
star mass \cite{Ott2012}, they performed their simulations starting
with the core of a presupernova star that had a mass of $12$ $M_\odot$
at zero-age main sequence.

\begin{figure}[t]
    \centerline{\includegraphics[width=8.6cm]{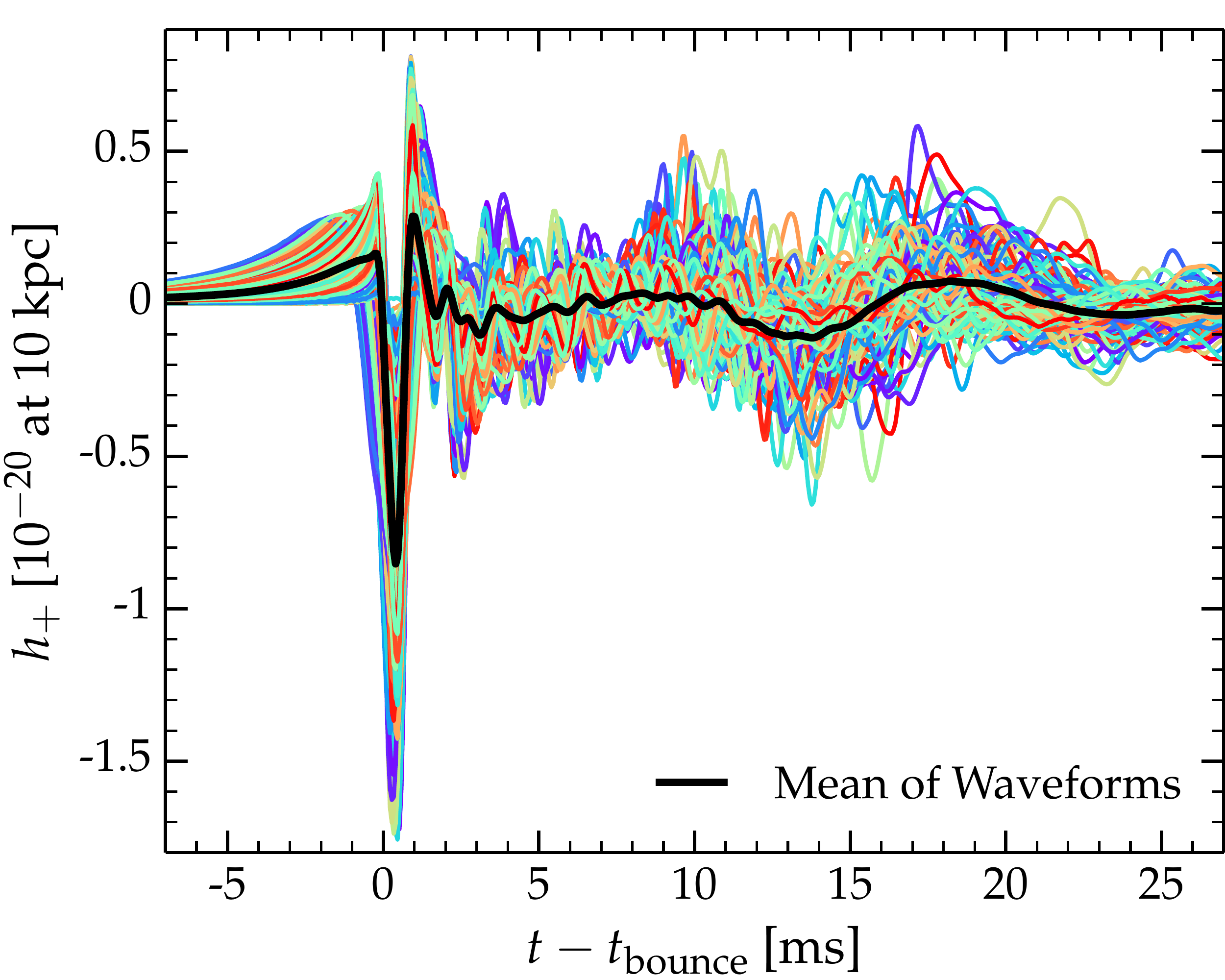}}
    \caption{\label{fig:meanWF} \small The 92 GW waveforms from the primary
Abdikamalov~\emph{et al.} catalog superimposed in varying colors.  The waveforms
are aligned to the point in time of core bounce and are resampled to have the
same sampling frequency.  The mean waveform of the catalog is overlaid in black.
 It is computed by taking the mean of the 92 waveforms at each point in time.}
\end{figure}

Abdikamalov~\emph{et al.}\ systematically varied the initial central
angular velocity $\Omega_{c}$ from $1\,\mathrm{rad\,s}^{-1}$ to
$15.5\,\mathrm{rad\,s}^{-1}$ and considered five different length
scales for differential rotation of $A1 = 300\,\mathrm{km}$, $A2 =
417\,\mathrm{km}$, $A3 = 634\,\mathrm{km}$, $A4 = 1268\,\mathrm{km}$,
and $A5 = 10000\,\mathrm{km}$ (see their Eq.~1). The
Abdikamalov~\emph{et al.}\ waveforms are split into a set of $92$
``catalog'' waveforms and a set of 43 ``injection'' waveforms. The
injection waveforms have one of the $A$ values listed in the above,
but values of $\Omega_{c}$ in between those covered by the catalog
waveforms. A small set of injection waveforms was calculated with a
different equation of state and with variations in the electron
capture prescription during collapse.  Abdikamalov~\emph{et al.} used
the injection waveforms to test their algorithms for extracting total
rotation and precollapse differential rotation from an observed
signal. In the present study, we primarily use the 92 catalog
waveforms and at times the subset of the injection waveforms that does
not include waveforms computed with different equation of state and
electron capture prescription. Figure~\ref{fig:meanWF} shows a
superposition of all 92 catalog waveforms (aligned to the time of core
bounce) and the mean waveform obtained by computing the average over
all waveforms.

While Abdikamalov~\emph{et al.}\ set up their models in the above way,
they point out that the initial angular velocity $\Omega_c$ is not a
good parameter to study: Progenitor cores with different structure
(e.g., less or more compact), but with the same $\Omega_c$ will lead
to different rotation rates at bounce, since, due to angular momentum
conservation, $\Omega$ increases $\propto r^{-2}$.  So an initially
further-out mass element (at greater initial $r$) will spin up more
than an initially further-in mass element at the same initial
$\Omega_c$. Abdikamalov~\emph{et al.} find that both the angular
momentum content of the inner core \emph{measured at bounce} and its
ratio of rotational kinetic energy to gravitational energy
$\beta_\mathrm{ic,b} = (T/|W|)_\mathrm{ic,b}$ are much more robust
parameters and are approximately independent of progenitor
structure~\cite{Ott2012}.  We note that the degree of precollapse
differential rotation is subject to very similar degeneracies as the
precollapse $\Omega_c$. A given fixed value of $A$ will lead to
different inner core rotation at bounce for different progenitor
structure, even if the total angular momentum inside the inner core is
the same. Hence, the results on differential rotation obtained by
Abdikamalov~\emph{et al.} are progenitor dependent (the strength of
this dependency remains to be established) and so will be the results
on differential rotation presented in this paper.

Another limitation of the Abdikamalov~\emph{et al.} study is the use
of only five discrete values of the differential rotation
parameter $A$, which is rather sparse and may not fully probe
the range of effects that variations in differential rotation may
have on rotating core collapse waveforms.

\subsection{Multivariate Regression Model: Overview}
\label{sec:MethodsOverview}

In the following sections, we describe in detail the methodology
required to construct a multivariate regression model for GWs from
rotating core collapse.  First, in~\ref{section:constructingthemodel},
we construct the baseline statistical model step by step.  In the
resulting matrix equation, the Fourier domain GW catalog waveforms are
simultaneously expressed as linear combinations of a yet unknown set
of feature vectors.  Each feature vector signifies an effect
contributed to the rotating core collapse GW signals associated with a
physical parameter.  In Sec.~\ref{section:encoding}, we
describe useful methods to encode representations of the physical
parameters of the progenitors into our statistical model.  Then in Sec.~\ref{section:basis}, we express the
feature vectors that characterize initial parameter effects themselves
as linear combinations of PCs, a set of orthonormal basis vectors.
This basis is derived using Singular Value Decomposition
(SVD)~\cite{Heng2009,Strang}.  The resulting statistical model is
given in Eq.~\ref{eq:finalmodel}.  Finally, we
provide the least squares solutions in~\ref{section:solutions} and discuss the use of statistical hypothesis testing in Sec.~\ref{sec:statistics}.

\subsection{Constructing the Statistical Model}
\label{section:constructingthemodel}

We begin by describing the preprocessing of the time domain GWs, and
then cast the statistical model in the frequency domain. In the time
domain, each waveform in the catalog is interpolated to have a
sampling frequency of 16384 Hz, Tukey windowed, and zero-padded. Then
they are aligned to core bounce, which is determined by the point in
time where the core has the highest central density. The aligned
waveforms are depicted in Fig.~\ref{fig:meanWF}.  The zero-padded ends
of the waveforms are then truncated so each is one second long.  Each
waveform is then Fourier transformed, and the real and imaginary parts
are kept unaltered.  In order to obtain a set of principal component
vectors (PCs), SVD is performed on the complex valued waveform
catalog~\cite{Heng2009, Strang}.  The role this basis plays in the
model is described in Sec.~\ref{section:basis}.  For the detector
noise model, we use the expected design-sensitivity zero-detuning
high-power Advanced LIGO noise \cite{LIGO-sens-2010}.

We describe the construction of the model in steps.  First we
construct a univariate version that considers just the $i$th waveform
in the catalog, a $1 \times t$ vector $\yv_{i}$, and its set of $p$
physical parameters, the $1 \times p$ vector $\xv_{i}$.  We then
expand the univariate equation into a full multivariate model,
considering all waveforms in the catalog simultaneously.  We describe
how physical parameters are encoded into each vector
$\xv_{i}$ in the univariate case and in the \emph{design matrix} $\X$,
in the multivariate case in Sec.~\ref{section:encoding}.

The $i$th waveform in the catalog is written as a linear combination
of unknown vectors arranged row-wise in $\M$,
\begin{equation}
\label{eq:basic}
\yv_{i} = \xv_{i} \M + \mathbf{r}_{i} \,,
\end{equation}
where $\M$ is a $p \times t$ matrix of $p$ unknown \emph{feature
  vectors}.  Each row vector, or feature vector, in $\M$ represents
the linear effect of a parameter value encoded in a column of the $1
\times p$ vector $\xv$.  We note that our use of term ``feature
vector'' is semantically different than it's use in the machine
learning literature. In Sec.~\ref{section:basis}, we will return to
$\M$ and discuss it in more detail.  The vectors $\yv_{i}$ and
$\xv_{i}$ are known and represent the $i$th waveform and the $i$th set
of initial conditions representing it, respectively.

Since some set of $p$ feature vectors in $\M$ is unlikely to provide a
perfect linear reconstruction of $\yv_{i}$, we include the vector
$\mathbf{r}_{i}$ as a residual error term.  This residual is due only
to the difference between the waveform $\yv_{i}$ and its linear model,
$\xv_{i}\M$.  If $\M$ could perfectly reconstruct all catalog
waveforms then that would mean that our linear model and parameter
encoding scheme was an exact predictor of waveform morphology for all
catalog waveforms.  Since core collapse is a highly complicated process,
we describe model uncertainty by assuming that this residual
is a complex multivariate normally distributed
random vector~\citep{Brillinger1981} with zero mean and a covariance matrix denoted by $\Sig_{R}$,
\begin{equation}
\label{eq:distr}
\underset{1 \times t}{\mathbf{r}_{i}} \sim \mathcal{N}^{C}(\underset{1 \times
t}{\mathbf{0}}, \underset{t \times t}{\Sig_{R}}) \,.
\end{equation}

We succinctly denote its multivariate normal probability distribution
using sampling notation~\cite{Marden}.
$\mathbf{v}~\sim~\mathcal{N}^{C}(\mathbf{a},\Sig)$ signifies a complex
multivariate normally distributed random vector $\mathbf{v}$ that is
parameterized by its central location, or expectation value,
$\mathbb{E}(\mathbf{v}) = \mathbf{a}$ and a positive-semidefinite covariance
matrix $\Sig$~\cite{Giri1977}.  Note that we assume throughout that the real and complex parts of our complex normal random vectors are independent (see Appendices of~\cite{Rover2009, Veitch2010}).  The $(i,j)$ element of a covariance matrix is
defined as the covariance between the $i$ and $j$ elements of the random vector $\mathbf{v}$.
Equivalently, we can write,
\begin{equation}
\Sigma_{i,j} = \mathbb{E}[(v_{i} - \mathbb{E}(v_{i}))(v_{j} - \mathbb{E}(v_{j}))^{\dagger}] \,.
\end{equation}
When helpful, we will underset the dimensions of quantities written in
matrix equations or written in sampling notation (where the $\sim$ is read as ``is sampled from'').  Throughout this
paper, we denote the conjugate transpose with $^\dagger$, and a
transpose of a real valued matrix with a superscript $^{T}$.

Each element of the diagonal of $\Sig_{R}$ in Eq.~\ref{eq:distr}
is then the covariance of the corresponding element of the vector
$\mathbf{r}_{i}$ with itself (the variance), and each off-diagonal element is
the covariance between the $i$th and $j$th elements of $\mathbf{r}_{i}$.
Assuming normality in the residuals is supported by the central limit theorem:
sums or products of random variables tend towards a Gaussian
distribution~\cite{Brillinger1981}, and a Gaussian distributed random vector (time domain signal) implies gaussianity of its Fourier Transform~\cite{Rover2009}.  If the normality
assumption is applicable, the mean vector and covariance matrix
completely characterize the random behavior of the system.

A model with increased uncertainty in the waveform due to GW detector noise is
of much greater interest.  We define $\yv'_{i} \equiv \yv_{i} + \sv_{i}$, where
$\sv_{i}$ is commonly approximated as a sample of additive, stationary, and
colored Gaussian noise from a given GW detector.  In the Fourier domain, the
detector noise is commonly assumed to be of Gaussian character with zero mean
and covariance matrix $\Sig_{S}$,

\begin{equation}
\label{eq:freqnoise1}
\underset{1 \times t}{\sv_{i}}  \sim \mathcal{N}^{C}(\underset{1 \times
t}{\mathbf{0}_{\vphantom{S}}} \,, \underset{t \times t}{\Sig_{S}}) \,.
\end{equation}
As commonly done in the GW data analysis community, we approximate
$\Sig_{S}$ as the zero matrix, but set its diagonal elements to the
variances of each frequency bin of the power spectral density (PSD)
that characterizes the noise of a given detector~\cite{Finn1992, Veitch2010}.  No
approximation need be made however, and a full noise covariance matrix
for a given detector could be used.

This allows us to rewrite Eq.~\ref{eq:basic} as,
\begin{equation}
\label{eq:twonoise}
\yv'_{i} = \xv_{i} \M + \mathbf{r}_{i} + \sv_{i}  \,.
\end{equation}
Since the sum of two normally distributed random variables is also normally
distributed~\cite{Marden, Brillinger1981}, we can combine the noise and error terms, setting $\mathbf{e}_{i} = \sv_{i} + \mathbf{r}_{i}$.  Equation~\ref{eq:twonoise} then becomes,
\begin{subequations}
    \begin{equation}
    \label{eq:ystar}
    \yv'_{i} = \xv_{i} \M + \mathbf{e}_{i} \,,
    \end{equation}
    \begin{equation}
    \label{eq:twonoisedist}
    \underset{1 \times t}{\mathbf{e}_{i}} \sim \mathcal{N}^{C}(\underset{1 \times t}{\mathbf{0}_{\vphantom{R}}} \,,
    \underset{t \times t}{\Sig_{R}} + \underset{t \times t}{\Sig_{S}}) \,.
    \end{equation}
\end{subequations}
From Eq.~\ref{eq:twonoise}, we can see that the distance of the source
(which sets the signal amplitude at the detector) determines the
degree to which instances of additive detector noise $\sv_{i}$ degrade
the signals.  Therefore, at the start of an analysis based on this
model, each $\yv_{i}$ needs to be scaled to a given source distance.

Up until this point, the structure of our statistical model is
identical to the model by R\"over~\emph{et al.}~\cite{Rover2009}.
Specifically, our Eq.~\ref{eq:ystar} is essentially identical to their
Eq.~6.  However, we consider the feature vectors in $\M$ to be unknown
quantities, and each $\xv_{i}$ known beforehand.  Past this point, we
depart from the methodology of~\cite{Rover2009}.

We form the multivariate analog of Eq.~\ref{eq:ystar} by including all
$n$ waveforms $\yv_{i}$ and all $n$ vectors $\xv_{i}$ into a matrix
equation.  Each $\yv'_{i}$ becomes a row in $\Y'$, each $\xv_{i}$
becomes a row in $\X$, and each $\mathbf{e}_{i}$ becomes a row in
$\mathbf{E}$.  The matrix of feature vectors $\M$ remains unchanged
when moving to the multivariate model --- different linear
combinations of the same feature vectors reconstruct different
waveforms.  We write the multivariate version of this model as,

\begin{subequations}
    \begin{equation}
    \label{eq:multivar}
	\underset{n \times t}{\mathbf{Y}^{'}} \hspace{2mm} = \hspace{2mm}
	\underset{n \times p}{\mathbf{X}} \hspace{3mm}
	\underset{p \times t}{\mathbf{M}}  +
	\underset{n \times t}{\mathbf{E}} \,,
    \end{equation}
    \begin{equation}
    \label{eq:modeldist}
    \mathbf{e}_{i} \sim \mathcal{N}^{C}(\mathbf{0}_{\vphantom{R}} \,,
    \Sig_{R} + \Sig_{S}) \,.
    \end{equation}
\end{subequations}
%
%

\subsection{Parameterizing The Design Matrix}
\label{section:encoding}

In this section, we summarize the methods we use for parameterizing
the \emph{design matrix} $\X$.  This is a crucial aspect of the
proposed multivariate regression model because the elements of
$\X$ define the linear combinations of the feature vectors in $\M$
that reconstruct the catalog signals. The description of the physical parameters within the design matrix determines the interpretation of the resulting feature vectors.

Information on any kind of initial condition, characteristic quantity, and simulation parameters can be incorporated, such as the rotation rate of the inner core at bounce ($\beta_{ic,b}$), the equation of state, the differential rotation profile ($A$), or the inner core electron fraction at bounce.

The translation of physical parameters into a meaningful design matrix
is known in the statistical literature as \emph{variable encoding}~(see, e.g., \cite{CohenCohen,Serlin1985}).  The variable encoding techniques described and applied in this paper are a small sample of many possible encoding schemes.

\subsubsection{Polynomial Encoding}
\label{sec:polyencoding}

In curve fitting, it is common to fit a curve to points in a
two-dimensional scatter plot using polynomials of some specified
order, allowing one to find evidence of trends in the data points.
This approach is also useful in our multivariate model.  For
instance, we can imagine that as the rotation rate at core
bounce changes, the presence of one of the feature vectors in the
catalog waveforms changes in a correlated fashion.

To encode polynomial functions of a physical parameter into the design
matrix, the actual values of the to-be-encoded physical parameter of
the $i$th waveform are placed in the $i$th row of $\X$.  The number of
columns in $\X$ devoted to encoding this parameter is equal to the
order of the polynomial being used.  In the first-order column, the
parameter values are unchanged.  In the second-order column, each of
the parameter values is squared.  In the third-order column, cubed,
and so on.  Each of these $\X$ columns is associated with a feature
vector in matrix $\M$.

Analogous to fitting a polynomial to a one dimensional curve, we fit a
polynomial function of the parameters, expressed by the feature
vectors in $\M$, to the set of waveforms $\Y$.  Also note that an
intercept term, or zeroth-order polynomial, is included.  This
manifests itself in the design matrix as a column in $\X$ where each
element is set to one.  We denote a column in $\X$ that is all ones as
$\mu$.

Each of the encodings described in this section includes a column of
ones, but how this column is interpreted depends on the encoding.  In
a polynomial encoding, a column of ones in the design matrix produces
a feature vector, $\mv_{\mu}$, that can be considered the constant
term of our polynomial function of the physical parameters.  Usually,
little attention is given to the morphology of the intercept feature
vector $\mv_{\mu}$, because $1 \cdot \mv_{\mu}$ is present in the
linear combination of feature vectors for every waveform
reconstruction (or waveform prediction).

To illustrate the polynomial encoding, we will use a
brief example.  Assume we have a catalog with three waveforms,
$\yv_1$, $\yv_2$, $\yv_3$, and that each waveform has a unique value
for some continuous parameter called $P$.  $\yv_1$ has parameter
$P_1$, $\yv_2$ has parameter $P_2$ and $\yv_3$ has parameter $P_3$.
We wish to see whether we can find feature vectors that follow, for
example, linear or quadratic trends in the waveforms.  We can write
out our second-order polynomial model, $\Y = \X_P \M$, explicitly,
\[
 \begin{pmatrix}
  \yv_1  \\
  \yv_2  \\
  \yv_3
 \end{pmatrix}
=
\bordermatrix{
 & \mu & linear & quadratic \cr
 & 1 & P_1 & P_1^2  \cr
 & 1 & P_2 & P_2^2  \cr
 & 1 & P_3 & P_3^2
                }
 \begin{pmatrix*}[l]
  &\mv_{\mu}   \\
  &\mv_{linear}   \\
  &\mv_{quadratic}
 \end{pmatrix*} \,\,.
\]

Later in Secs.~\ref{section:basis} and~\ref{section:solutions}, we use
least squares to solve for the matrix of feature vectors $\M$ as a
linear combination of PCs.

While our multivariate regression model is linear in the sense that
catalog waveforms are constructed by linear combinations of feature
vectors, non-linear functions of the physical parameters can be used
to produce those feature vectors.  This allows for great flexibility
in modeling the influence of physical parameters on rotating core
collapse waveforms.  Besides polynomials, other basis functions can be
used, such as splines or radial basis functions~\cite{Tibshirani}.

Some parameters used to specify initial conditions for rotating core
collapse are difficult to model continuously.  For example, only five
differential rotation profiles were employed by Abdikamalov~\emph{et
  al.}~\cite{Abdikamalov2013}.  Polynomials may not be the most
suitable encoding.  Also, it may be desirable to partition a parameter
into several bins in order to see if there are particular feature
vectors associated with, for instance, ``low'', ``medium'', or
``high'' parameter values.  The following two types of variable
encoding are devoted to discrete parameters.  For example,
Abdikamalov~\emph{et al.}, simulated the core collapse of progenitors
where each had one of five differential rotation profiles.

\subsubsection{Deviation Encoding}
\label{sec:devencoding}
  
It is more straightforward to illustrate, instead of describe, a
deviation encoding of the design matrix $\X$.  For example, say we
wish to partition a six-waveform catalog into three groups, defined by
some physical parameter that takes on three values (or three ranges of
values).  Under a deviation encoding, waveforms in these groups
(labeled by the subscripts $g_1$, $g_2$ and $g_3$) are represented
using three feature vectors; one for the mean of all catalog
waveforms, labeled $\mv_{\mu}$; one for the average difference from
the mean of waveforms in $g_1$, labeled $\mv_{g_1 - \mu}$; and one for
the average difference of waveforms in $g_2$, labeled $\mv_{g_2 -
  \mu}$.  The average difference from the mean of $g_3$ waveforms is
given by the negative of the sum of the $g_1$ and $g_2$ differences.
We illustrate this encoding assuming there are a total of six
waveforms in the catalog, two from each of the three groups.  We write
out this instance of $\Y = \X \M$ as,
\[
 \begin{pmatrix*}[l]
  &\yv_{1(g_1)}   \\
  &\yv_{2(g_1)}   \\
  &\yv_{3(g_2)}   \\
  &\yv_{4(g_2)}   \\
  &\yv_{5(g_3)}   \\
  &\yv_{6(g_3)}
 \end{pmatrix*}
=
\bordermatrix{
 & \mu & g_1 - \mu & g_2 - \mu \cr
 & 1 & 1 & 0  \cr
 & 1 & 1 & 0  \cr
 & 1 & 0 & 1  \cr
 & 1 & 0 & 1  \cr
 & 1 &-1 &-1  \cr
 & 1 &-1 &-1
                }
 \begin{pmatrix*}[l]
  &\mv_{\mu}   \\
  &\mv_{g_1 - \mu}   \\
  &\mv_{g_2 - \mu}
 \end{pmatrix*} \,.
\]
Throughout the paper, we refer to the columns of $\X$, except the
intercept term ($\mu$), as \emph{comparisons}.  For instance, we can say
that the second column of $\X$, $g_1 - \mu$, is a comparison between
the mean of the $g_1$ waveforms and the mean of all six waveforms.  If
the mean of the $g_1$ waveforms is the same (or very similar) to the
mean of all six waveforms, then the $\mv_{g_1 - \mu}$ feature vector
will be insubstantial, or insignificant --- many of the elements of
$\mv_{g_1 - \mu}$ will be zero or very close to zero.  This deviation
encoding pattern is extensible to any number of groups, and any number
of catalog waveforms.

\subsubsection{Dummy Variable Encoding}
\label{sec:dummyvarencoding}

A variation of deviation encoding expresses catalog waveforms as a
difference from a specified reference group, instead of as a
difference from the mean of the whole catalog.  The name ``dummy
variable'' refers to using ones as logical placeholders for actual
parameter values in the design matrix~\cite{CohenCohen}.  Using the
same notation used previously, we designate the reference group in the
next example to be $g_1$.  In the following case, each group is
described as its difference from the average of the $g_1$ waveforms,
instead of by its difference from the catalog mean.  Explicitly, this is written as,
\[
 \begin{pmatrix*}[l]
  &\yv_{1(g_1)}   \\
  &\yv_{2(g_1)}   \\
  &\yv_{3(g_2)}   \\
  &\yv_{4(g_2)}   \\
  &\yv_{5(g_3)}   \\
  &\yv_{6(g_3)}
 \end{pmatrix*}
=
\bordermatrix{
 & \mu & g_2 - g_1 & g_3 - g_1 \cr
 & 1 & 0 & 0  \cr
 & 1 & 0 & 0  \cr
 & 1 & 1 & 0  \cr
 & 1 & 1 & 0  \cr
 & 1 & 0 & 1  \cr
 & 1 & 0 & 1
                }
 \begin{pmatrix*}[l]
  &\mv_{\mu}   \\
  &\mv_{g_2 - g_1}   \\
  &\mv_{g_3 - g_1}
 \end{pmatrix*} \,,
\]

The first column, $\mu$, is the intercept term.  In this dummy
variable encoding, $\mv_{\mu}$, is the mean of the $g_1$ waveforms.
The second column, $g_2 - g_1$, is a comparison of the mean of the
$g_1$ group to the mean of the $g_2$ group.  The feature vector
$\mv_{g_2 - g_1}$ is therefore the difference between the mean of the
$g_2$ and the $g_1$ waveforms.  The third column, the $g_3 - g_2$
comparison, along with its feature vector, $\mv_{g_3 - g_1}$, is
interpreted in a similar fashion. 
Linear combinations of the feature
vectors determined by the design matrix reconstruct the six waveforms
as 
\[
 \begin{pmatrix*}[l]
  &\yv_{1(g_1)}   \\
  &\yv_{2(g_1)}   \\
  &\yv_{3(g_2)}   \\
  &\yv_{4(g_2)}   \\
  &\yv_{5(g_3)}   \\
  &\yv_{6(g_3)}
 \end{pmatrix*} =
 \begin{pmatrix*}[l]
  \mv_{\mu}          \\
  \mv_{\mu}          \\
  \mv_{\mu} + \mv_{g_2 - g_1}           \\
  \mv_{\mu} + \mv_{g_2 - g_1}           \\
  \mv_{\mu} - \mv_{g_3 - g_1}           \\
  \mv_{\mu} - \mv_{g_3 - g_1}
 \end{pmatrix*}
\,.
\]
As before, the ${g_1}$ subscript labels waveforms that are considered
members of the $g_1$ group, and so on.  As with the deviation
encoding, this same encoding pattern is extensible to any number of
waveform groups and any number of catalog waveforms.

\subsubsection{Multiple Parameters and Interactions}
\label{sec:multipleparam+interactions}

Generally, more than one physical parameter is varied in core collapse
simulations.  As an example, imagine that we can partition our six
waveforms as belonging to one of three groups, $g_1$, $g_2$ or $g_3$,
as before. Additionally, the same set of waveforms can also be
partitioned into one of two other groups, labeled $h_1$ and $h_2$.
For example, the three groups $g_1$, $g_2$ and $g_3$, might represent
the fact that these waveforms were produced from progenitors with
differential rotation $A1$, $A2$, and $A3$, respectively.  The
waveforms in groups $h_1$ and $h_2$ may then have come from
progenitors with two different equations of state.  Using a
hypothetical waveform catalog with six waveforms as before, with two
waveforms in each of the $g$ groups and three waveforms in each of the
$h$ groups, we can construct a joint design matrix for both
parameters.

To illustrate, we use the same deviation encoding on $g$ shown in
Sec.~\ref{sec:devencoding}, and then choose a dummy variable encoding
on $h$, where $\yv_1$, $\yv_2$ and $\yv_3$ are members of $h_1$, and
the other three waveforms are members of $h_2$.  We choose our
reference group to be $h_2$.  This design matrix, $\X_{g,h}$ is
written explicitly as,
\[ \X_{g,h} =
\bordermatrix{
 & \mu & g_1 - \mu & g_2 - \mu & h_2 - h_1 \cr
 & 1 & 1 & 0 & 0 \cr
 & 1 & 1 & 0 & 0 \cr
 & 1 & 0 & 1 & 0 \cr
 & 1 & 0 & 1 & 1 \cr
 & 1 &-1 &-1 & 1 \cr
 & 1 &-1 &-1 & 1
                }         \,.
\]
Concatenating the encodings of different physical parameters
(i.e.\ multiple groups) into the same design matrix allows us to
consider the dependence of a waveform's morphology on different
physical parameters as a linear combination of feature vectors, each
attributable to one of the parameters.  To help illustrate this subtle
but important point, we write out explicitly how the feature vectors
produced by the above design matrix construct the six example catalog
waveforms,
\[
 \begin{pmatrix*}[l]
  &\yv_{1(g_1,h_1)}   \\
  &\yv_{2(g_1,h_1)}   \\
  &\yv_{3(g_2,h_1)}   \\
  &\yv_{4(g_2,h_2)}   \\
  &\yv_{5(g_3,h_2)}   \\
  &\yv_{6(g_3,h_2)}
 \end{pmatrix*} =
 \begin{pmatrix*}[l]
  &\mv_{\mu} + \mv_{g_1 - \mu}  \\
  &\mv_{\mu} + \mv_{g_1 - \mu}  \\
  &\mv_{\mu} + \mv_{g_2 - \mu}  \\
  &\mv_{\mu} + \mv_{g_2 - \mu} + \mv_{h_2 - h_1} \\
  &\mv_{\mu} - \mv_{g_1 - \mu} + \mv_{g_2 - \mu} + \mv_{h_2 - h_1}  \\
  &\mv_{\mu} - \mv_{g_1 - \mu} + \mv_{g_2 - \mu} + \mv_{h_2 - h_1}  \\
 \end{pmatrix*} \,\,.
\]
Once two encodings of two (or more) parameters, or groups, have been
concatenated into the same design matrix, the interpretation of the
feature vectors changes.  For example, the feature vector $\mv_{g_{1}
  - \mu}$ is now interpreted as the average difference from the
catalog mean of the waveforms in the $g_1$ group \emph{after the
  removal of waveform morphology correlated with waveforms in either
  of the $h$ groups}.  Note also that in this example, $\mv_{\mu}$
cannot be both the average of all catalog waveforms and the average of
the waveforms in the $h_1$ group.  It's precise physical meaning is
difficult to qualify, especially as the complexity of the design
matrix grows.  It is best referred to as the ``intercept feature
vector''.

In some cases, it may be desirable to consider \emph{interactions}
between groups, where an interaction defines the set of catalog
waveforms that are members of multiple groups.  For instance, we may
be interested in features present only in waveforms that are
considered members of one group \emph{and} of a second group. Using
the above example, we can produce feature vectors unique to waveforms
in both $g_1$ and $h_1$, and $g_2$ and $h_1$, where we use the
$\times$ symbol to denote an interaction between two groups,
\[
\X_{g,h,g \times h} =
\]
\[
\bordermatrix{
 & \mu & g_1 - \mu & g_2 - \mu & h_2 - h_1 & g_1 \times h_1 & g_2 \times h_1 \cr
 & 1 & 1 & 0 & 1 & 1 & 0 \cr
 & 1 & 1 & 0 & 1 & 1 & 0 \cr
 & 1 & 0 & 1 & 1 & 0  & 1 \cr
 & 1 & 0 & 1 & 0 & 0 & 0 \cr
 & 1 &-1 &-1 & 0 & 0 & 0 \cr
 & 1 &-1 &-1 & 0 & 0 & 0
                }  \,.
\]
An interaction column is computed easily by an element-wise
multiplication of two columns in the design matrix~\cite{CohenCohen}.
A design matrix with a polynomial encoding can be concatenated with a
design matrix with a dummy variable encoding, and interactions between
a polynomial encoded independent variable and a deviation encoded
variable are computed by an element-wise multiplication of design
matrix columns.  These two rules for producing interaction terms and
modeling multiple groups concurrently applies to all encoding
types~\cite{CohenCohen}.  In the above illustration, we created what is called a \emph{two-way interaction} between two different parameter types.  By multiplying more than two design matrix columns together at a time, higher order interactions terms can be defined.

\subsection{Factoring $\M$ with Singular Value Decomposition}
\label{section:basis}

In the previous sections, $\M$ is treated as an unknown matrix of
physically meaningful feature vectors which can be used to reconstruct
each of the waveforms $\yv_{i}$.  At this point, we can estimate the
$p \cdot t$ matrix elements in $\M$ by solving the matrix equation $\Y
= \X \M$ using least squares.  For convenience, $p$ is the number of columns in $\X$, $k$ is the number of PCs in $\Zt$, and $t$ is the number of samples per waveform in $\Y$.

However, reducing the number of statistical parameters
  (elements of $\M$) that need to be estimated greatly reduces the
  degrees of freedom and enables the apparatus of statistical
  hypothesis testing (see Sec.~\ref{sec:statistics} for further
  details on hypothesis testing).  To reduce the number of matrix
  elements that need to be estimated, we factor $\M$ into two matrices
  in such a way that our feature vectors are expressed as linear
  combinations of PCs.  Given a PC basis, this unknown matrix is
  comprised of $p \cdot k$ PC coefficients, where $p \cdot k \ll p
  \cdot t$.  Refs.~\cite{Heng2009,Logue2012} have shown that for $n$
  rotating core collapse waveforms, only $k \ll n$ basis vectors are
  needed to provide excellent reconstructions of a large majority of
  waveforms of the catalog.

\begin{figure}[t]
    \centerline{\includegraphics[width=8.6cm]{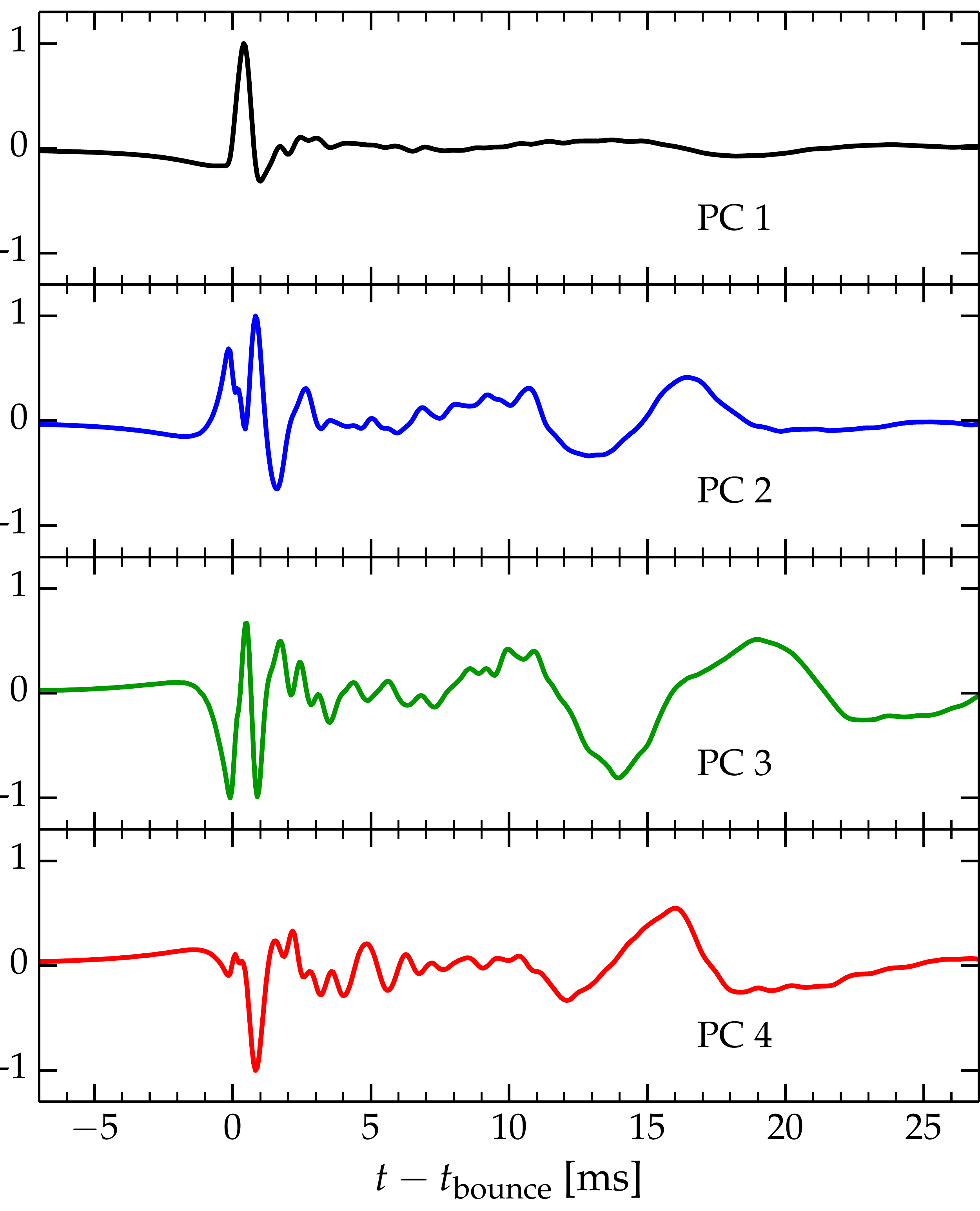}}
    \caption{\small The first four principal components (PCs) from the
      waveforms of the Abdikamalov~\emph{et al.} catalog in the time
      domain.  Each PC has been normalized by its maximum amplitude.}
    \label{fig:svd}
\end{figure}

To construct the PC basis, we follow previous work
\cite{Heng2009,Logue2012,Rover2009} and apply singular-value
decomposition (SVD) to factorize our matrix of Fourier-transformed
waveforms, $\Y$, into three matrices,
\begin{equation}
\Y = \mathbf{U} \mathbf{S} \mathbf{V^{\dagger}} \,,
\end{equation}
where the rows of $\mathbf{V^{\dagger}}$ are the eigenvectors of the
matrix $\Y^\mathbf{\dagger} \Y$ and are called principal components
(PCs), which form an orthonormal basis for $\Y$.  The PCs obtained in
this fashion are equivalent to those obtained by applying SVD to the
time domain waveforms, Fourier transforming the time domain PCs, then
normalizing the PCs with the multiplicative constant
$t_{s}^{-\nicefrac{1}{2}}$, where $t_s$ is the number of time samples
per time domain waveform.  Figure~\ref{fig:svd} depicts the first four
PCs computed from the Abdikamalov~\emph{et al.}
catalog~\cite{Abdikamalov2013}.

Past work~\cite{Heng2009,Rover2009,Logue2012,Cannon2011} used SVD in
the following fashion to form a basis from which GWs are
reconstructed: To be exact, the $i$th catalog waveform is represented
as a linear combination of $k$ basis vectors.  We denote the $1 \times
k$ vector of coefficients of this linear combination by $\mathbf{a}$,
and the PC basis by $\mathbf{Z}$, whose columns are the first $k$ PCs.
Each $\yv_{i}$ is approximated by,
\begin{equation}
\yv_{i} \approx \sum_{j = 1}^{k} a_{j} \mathbf{Z}_{j}\,\,,
\end{equation}
where $\mathbf{Z}_{j}$ is the $j$th basis vector of the PC basis
$\mathbf{Z}$ and $a_{j}$ is the corresponding reconstruction
coefficient.

Instead of directly representing catalog waveforms with linear
combinations of PCs, our multivariate regression model represents the feature vectors that characterize physical parameters as linear combinations of PCs.  Subsequently, catalog waveforms are represented by linear combinations of these feature vectors, where each feature vector is a row in $\M$.  To express this relationship between the catalog waveforms and the PC basis, we factor $\M$ into a known and an unknown part,
\begin{equation}
\label{eq:M}
	\underset{p \times t}{\M} \hspace{2mm} = \hspace{2mm}
	\underset{p \times k}{\B} \hspace{3mm}
	\underset{k \times t}{\Zt} \,.
\end{equation}
where the rows of $\Zt$ are the $k$ PCs.  Since all other matrices,
$\Y$, $\X$, and $\Zt$, are known, what remains is to find a solution
for the $p \times k$ elements of $\B$, which we will obtain below via
a least-squares fit.

Casting our feature vectors as linear combinations of PCs is
beneficial in two ways.  First, we bridge between the past work
of~\cite{Heng2009,Rover2009,Logue2012} to the physical parameters of
collapse, whose relationship to GW morphology is of great interest.
Second, using the PC basis enables the apparatus of statistical
hypothesis testing by dramatically reducing the number of statistical
parameters that need to
be estimated (see Sec.~\ref{sec:statistics}).  Test statistics and
hypothesis testing can be used to measure the magnitude of a feature
vector associated with a physical parameter.

After the feature matrix $\M$ has been factored into $\B$ and $\Zt$,
we rewrite Eq.~\ref{eq:multivar} with $\mathbf{E} = \begin{bmatrix*}
  \ev_{1}^{T} & \ev_{2}^{T} & \ldots & \ev_{n} \end{bmatrix*}^T$ as
\begin{equation}
	\label{eq:finalmodel}
	\underset{n \times t}{\mathbf{Y}^{'}} \hspace{2mm} = 			\hspace{2mm}
	\underset{n \times p}{\mathbf{X}} \hspace{3mm}
	\underset{p \times k}{\mathbf{B}} \hspace{3mm}
	\underset{k \times t}{\mathbf{Z}^{\dagger}} +
	\underset{n \times t}{\mathbf{E}} \,.
\end{equation}
We note here that it is equivalent to speak about rows of $\B$ or rows of
$\M$ for referring to feature vectors associated with physical
parameters because each row of $\B$ defines the linear combination of
PC basis vectors that construct the corresponding feature vector in
$\M$.

\subsection{The Least Squares Solution}
\label{section:solutions}

With all the ingredients that are required to specify our linear model
at hand, we can move to estimating the unknown quantities in
Eq.~\ref{eq:finalmodel}, $\B$ and $\Sig_R$.  We denote estimators for
the unknown quantities with a caret ($\,\hat{\,}\,$), while the
\emph{true} value of an unknown quantity has the same bold notation as
known vectors and matrices.  In this section, we provide the known
analytic solutions for these estimators, which maximize the complex
multivariate Gaussian likelihood function over the
residuals~\cite{Marden, Giri1977}.  Maximizing this likelihood
function is equivalent to minimizing the sum of squares of the
elements of the residuals $\mathbf{R}$, where $\mathbf{R} = \Y - \X
\Best \Zt$.  In other words, our estimate of $\B$, denoted $\Best$,
minimizes the quantity,
\begin{equation}
\label{eq:minimization}
|| \Y' - \X \B \Zt ||^{2} \,,
\end{equation}
where from Eq.~\ref{eq:twonoise}, each $\yv_{i}' = \yv_{i} + \sv_{i}$.
The estimate of $\B$ which minimizes the above expression is given
analytically~\cite{Marden,Giri1977},
\begin{equation}
\label{eq:estimator}
\mathbf{\hat{B}}  = (\X^{T}\X)^{-1} \mathbf{X}^{T}\mathbf{Y'}\mathbf{Z}(\mathbf{Z}^{\dagger}\mathbf{Z})^{-1} \,.
\end{equation}

Equation~\ref{eq:estimator} can be simplified in two ways.  Since the
PCs produced from the SVD form an orthonormal basis set,
$\mathbf{Z}^{\dagger}\mathbf{Z}~=~\mathbf{I}_{t}$, the $t \times t$
identity matrix, where $t$ is the number of data samples in each of
the waveforms.  We can also factor the least squares solution for $\B$
into two parts, remembering that each $\yv'_{i} = \yv_{i} + \sv_{i}$.
This factored least squares estimator is written as,
\begin{equation}
\label{ls2}
\mathbf{\hat{B}} = \Cx \mathbf{X}^{T}\mathbf{Y}\mathbf{Z} +
					    \Cx\mathbf{X}^{T}
	\begin{bmatrix*} \sv_{1}^{T} & \sv_{2}^{T} & \ldots & \sv_{n} \end{bmatrix*}^T
					    \mathbf{Z} \,,
\end{equation}
where $\Cx = (\X^{T}\X)^{-1}$.  Instances of detector noise $\sv_{i}$
are unrelated to the model residual $\R$, and from
Eq.~\ref{eq:freqnoise1}, each of their expectation values is the zero
vector ($\mathbb{E}(\sv_{i}) = \mathbf{0}$).  Therefore, we can drop
the detector noise contribution to the estimator and set $\Y^{'} =
\Y$.  Equation~\ref{eq:estimator} simplifies to
\begin{equation}
\label{eq:ls}
\underset{p \times k}{\mathbf{\hat{B}}} = \Cx\mathbf{X}^{T}\mathbf{Y}  \mathbf{Z} \,,
\end{equation}
where $p$ is the number of columns of $\X$, and $k$ is the number of
PCs in $\Zt$.  Now that we have an estimate $\Best$ for $\B$, we can use our
multivariate regression model to generate waveforms with arbitrary
values of the physical parameters determined by our choice of the
design matrix $\X$.

To obtain \emph{reconstructions} of the catalog waveforms $\Y$, we can write,
\begin{equation}
\Y^{R} = \X \Best \Zt
\end{equation}
where the reconstructed waveforms are denoted $\Y^{R}$.  To
\emph{predict} a waveform from a progenitor with different parameter
values than any of the original catalog waveforms, we encode its
physical parameters into a vector $\mathbf{\tilde{x}}$ in the same
fashion as the original $\X$ was encoded and write,
\begin{equation}
\label{eq:predictioneq}
\mathbf{\tilde{y}} = \mathbf{\tilde{x}} \Best \Zt
\end{equation}
where $\mathbf{\tilde{y}}$ is the expected waveform predicted from our regression model.  In Eq.~\ref{eq:predictioneq}, $\X$, $\Best$ and $\Zt$ are derived from the original waveform set.

We can also use our regression model to examine how influential
certain physical parameters are on catalog morphology.  In
Sec.~\ref{section:encoding}, we saw how our encodings of the design
matrix led to $\B \Zt$ being interpretable as a feature matrix $\M$,
where each of the feature vectors in $\M$ is associated with a column
of the design matrix $\X$.  If the comparison defined by the $i$th
column of $\X$ is insignificant to waveform morphology, then we would
expect the magnitude of the $i$th feature vector in $\M$ to be small.
For the feature vector to have a small magnitude, the elements in the
$i$th row of $\B$ must be zero or close to zero.  Therefore, we can
measure how important various parameters are to catalog morphology by
looking closely at the magnitude of the elements of our estimator of
$\B$.  In the following section, we give test statistics based on the
values of $\Best$ that are useful for measuring how influential
particular physical parameters are on catalog morphology.

\subsection{Statistical Hypothesis Testing}
\label{sec:statistics}

In a statistical hypothesis test, two hypotheses are proposed, a null
hypothesis and its alternative hypothesis~\cite{PDG2012}.  In our situation, they can be summarized as follows:
\begin{itemize}
	\item Null Hypothesis, $H_0$: Relevant elements of $\B = 0$;

	\item The Alternative, $H_a$: Relevant elements of $\B \neq 0$.

\end{itemize}
In this paper, we are primarily interested in whether specific feature
vectors (rows of $\B$), are equal to the zero vector.  In this case,
our $H_0$ is that all the elements in a particular row of $\B$ are
equal to zero.  Occasionally, we may be interested in whether one of
the PC basis vectors is influential in a given feature.  In that case,
our $H_0$ is that a particular element of $\B$ is equal to zero.  We
describe in detail the procedure for conducting hypothesis tests on
the rows of $\B$ in Sec.~\ref{sec:hotellingt2}.  The procedure for
testing individual elements is given in Sec.~\ref{sec:studentt}.

\subsubsection{An Illustration}
\label{sec:illustration}

The evidence in favor of, or against, some null hypothesis ($H_0$)
depends not just on the magnitudes of the elements of $\B$ in
question, but also on the covariances of the waveforms.  Additionally,
the number of waveforms also plays a role.  As a simple example,
imagine we have put a dummy variable encoding on a set of waveforms
whose parameters can be grouped into three groups labeled $g_1$,
$g_2$, and $g_3$.  We are interested in whether there is a significant
difference between the $g_2$ and $g_1$ waveforms.  This is the
scenario described in Sec.~\ref{sec:dummyvarencoding}.

In this scenario, the feature vector $\mv_{g_2 - g_1}$ produced from
the design matrix is the average of the differences between the $g_2$
and the $g_1$ waveforms.  Our $H_0$ is that the elements in this row
of $\B$, the PC coefficients that construct the feature vector
$\mv_{g_2 - g_1}$, are all equal to zero --- there is no difference,
on average, between the $g_2$ and $g_1$ waveforms.  Imagine we find
that the magnitudes of these PC coefficients are somewhat large,
leading to a substantial feature vector $\mv_{g_2 - g_1}$.  This
result provides evidence against $H_0$.

However, if the morphology of this set of $g_2$ and $g_1$ waveforms is
very heterogeneous, then our evidence against $H_0$ diminishes.
Noting a large difference between two sets of highly variable
waveforms is less compelling than if the waveforms within each of the
two sets were very similar to each other.  We
construct the covariance matrix for the residuals below
in Sec.~\ref{sec:buildingsigr}. 

The number of $g_1$ or $g_2$ waveforms generated also matters.
Imagine we obtain a substantial feature vector, and the morphology of
the two sets of waveforms is reasonably homogeneous.  However, if there
were only two $g_2$ and two $g_1$ waveforms, it is less reasonable to
claim that $g_2$ and $g_1$ waveforms are significantly different than
if there were 20 $g_2$ and 20 $g_1$ waveforms.  This type of
information is captured by the inverse of the covariance matrix of the
design matrix, $\Cx = (\X^{T} \X)^{-1}$, which factors into the test
statistics.

\begin{figure}[t]
    \centerline{\includegraphics[width=8.6cm]{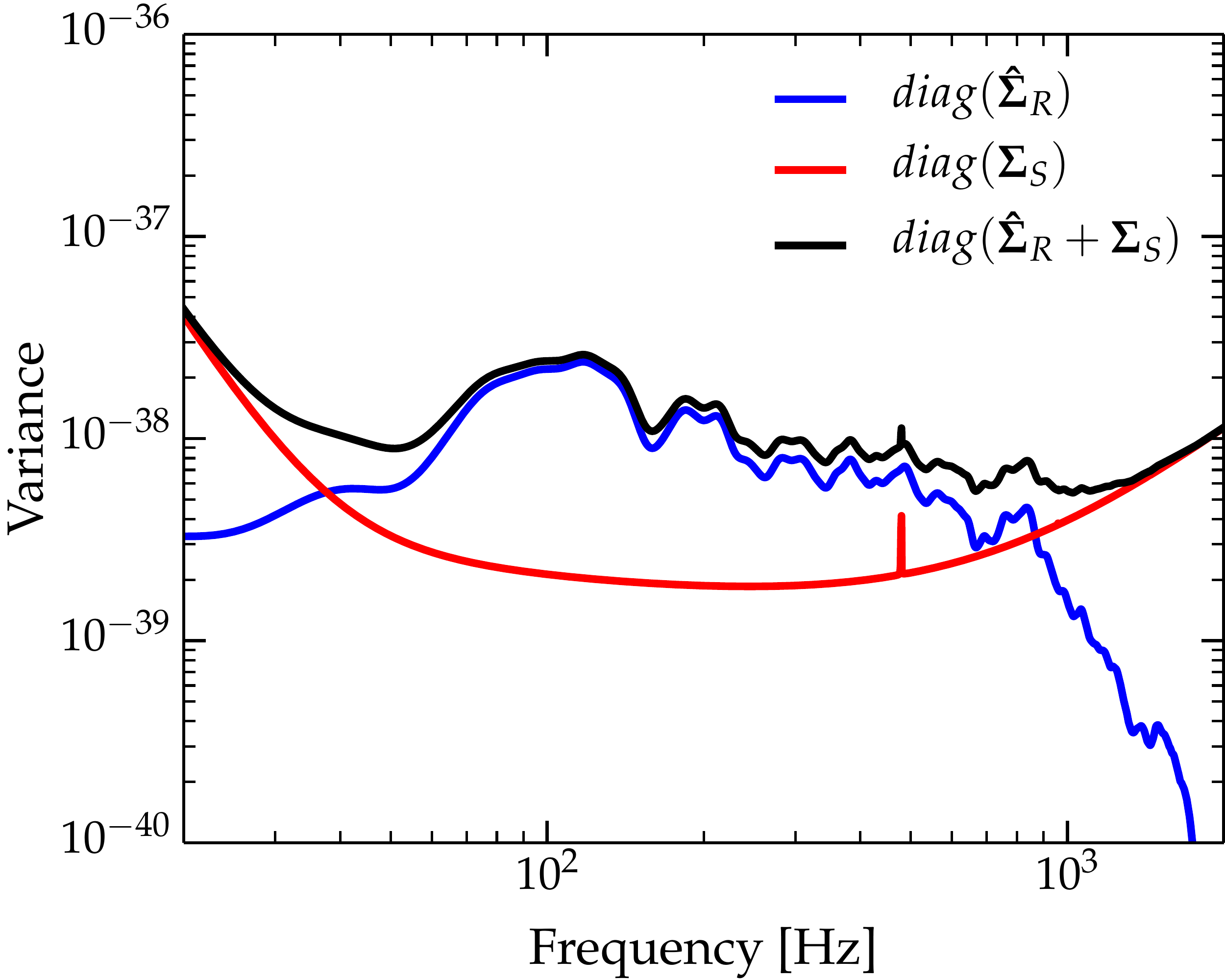}}
    \caption{\label{fig:RvsS} \small The diagonal of $\hat{\Sig}_{R}$, $\Sig_S$,
      and the sum of $\hat{\Sig}_R$ and $\Sig_S$.  We set the diagonal
      elements of $\Sig_S$ to the Advanced LIGO noise variances.  In
      producing $\hat{\Sig}_{R}$, the catalog waveforms have been
      scaled to a distance of $10\,\mathrm{kpc}$, and we used a
      design matrix with a deviation encoding on the 5 differential
      rotation profiles.  As the waveforms are scaled to greater
      distances, the noise curve variances will begin to dominate over
      the residual variances.}
\end{figure}

\subsubsection{Estimating the Covariance of the Residuals}
\label{sec:buildingsigr}

We express the level of heterogeneity of the morphology of a set of
waveforms with a covariance matrix on the residuals of our fit and the
original catalog waveforms.  The matrix of residuals, $\R$, can be
computed by,
\begin{equation}
\label{eq:residualmat}
\R = \Y - \X \Best \Zt \,.
\end{equation}
From~\cite{Marden,Giri1977}, we obtain an estimator for the covariance of the residuals, $\Sig_R$, as
\begin{equation}
\label{eq:covR}
\hat{\Sig}_{R} = \frac{1}{n - p} \mathbf{R}^{\dagger} \mathbf{R} \,,
\end{equation}
where $n$ is the number of catalog waveforms, and $p$ is the number of columns of $\X$.

We also want to include uncertainty due to detector noise in our inferences.  From Eq.~\ref{eq:twonoisedist}, we can add the detector noise covariance matrix (described in Eq.~\ref{eq:freqnoise1}) to obtain our estimate of the total error covariance, due to the combined hypothetical detector noise and the residuals, $\hat{\Sig}_{E}$,
\begin{equation}
\hat{\Sig}_{E} = \hat{\Sig}_R + \hat{\Sig}_{S} \,.
\end{equation}
In Fig.~\ref{fig:RvsS}, we graphically compare the diagonals of
$\hat{\Sig}_R$ and $\Sig_{S}$.  To produce this plot, we used a design
matrix with a deviation encoding on the five values of differential
rotation.  At a common source distance of 10 kpc, the variance due to
the residuals remains dominant over the variances due to
the Advanced LIGO design noise curve in the zero-detuning, high-power
configuration~\cite{LIGO-sens-2010}.

While the elements of our solution $\B$ are PC
coefficients, the elements of $\hat{\Sig}_R$ are the residual variance
and covariances between residual frequency bins.  We change the basis
of $\hat{\Sig}_R$ into the same PC basis as our solution $\B$ in order
to estimate the total error covariance in our test
statistics~\cite{Marden}, 
\begin{equation}
\label{eq:prop}
\mathbf{\hat{\Sigma}}_{Z} = \mathbf{Z}^{\dagger} \mathbf{\hat{\Sigma}}_{E} \mathbf{Z}  \,,
\end{equation}
where the total error covariance in terms of the PC basis is
$\hat{\Sig}_{Z}$.  We use this result in the construction of both
Hotelling's $T^2$ and student's $t$ test statistics.

\subsubsection{Hotelling's $T^2$ --- Inferences Regarding Rows of $\B$}
\label{sec:hotellingt2}

We are often interested in whether all the elements in a specific row
of $\Best$ are equal to zero.  This is because each row of $\Best$
determines how influential to catalog morphology each column of the
design matrix is.   We use the variable $\hat{\bv}_{i}$ to represent a selected row.  This particular test statistic is known as the Hotelling's $T^{2}$ statistic~\cite{Hotelling1931}, and is given by,
\begin{equation}
\label{eq:Hotelling}
T^{2} = \frac{ \hat{\bv}_{i} \mathbf{\hat{\Sigma}}_{Z}^{-1} \hat{\bv}_{i}^{\dagger} }{\Cx_{ii}}\,\, ,
\end{equation}
where $\Cx_{ii}$ is the $i$th diagonal element of $\Cx=~(\X^{T}
\X)^{-1}$.  The matrix $\Cx$
contains information regarding the number of waveforms, as per the
discussion in Sec.~\ref{sec:illustration}.  Under $H_0$ (all
elements in $\bv_{i} = \mathbf{0}$), it can be shown that this
statistic can be written in terms of the $\mathcal{F}$-distribution~\cite{Marden,Giri1977},
\begin{equation}
\label{eq:FtoP}
\frac{v - k + 1}{vk} T^{2} \sim \mathcal{F}_{2k,2(v-k+1)} \,,
\end{equation}
where $v = n - p$, $n$ is the number of waveforms in $\Y$, $p$ is the
number of columns of $\X$, and $k$ is the number of PCs in $\Zt$.  The
tilde ($\sim$) can be read as ``is distributed as''.  $2k$ is the
``upper'' degrees of freedom in the $\mathcal{F}$ distribution~\cite{JamesF}, and
$2(v - k + 1)$ is the ``lower'' degrees of freedom.  We delay a brief
discussion of the details and use of these test statistics until
Sec.~\ref{sec:teststatdiscussion}

Hotelling's $T^2$ statistic is valid if and only if $v \geq k$,
necessitating the use of our PC basis $\Zt$ in the statistical model
(see Sec.~\ref{section:basis}). If there were no basis used (i.e.,
$\Zt$ is set to the $t \times t$ identity matrix), then $k = t$ in
Eq.~\ref{eq:FtoP}, where $t$ is the number of data samples in each
waveform, $p$ is the number of design matrix columns and $k$ is the
number of PCs in $\Zt$.  In this case, $v = n - p$ is not greater than
or equal to $k$, causing the left hand side of Eq.~\ref{eq:FtoP} to be
negative --- outside the domain of the $\mathcal{F}$-distribution. The
constraint $v \geq k$ cannot be satisfied unless the waveforms are
reconstructed with a basis that is smaller than the size of the
catalog.  Thus using a PC basis not only allows us to connect PCs to
physical parameters, but also enables statistical hypothesis testing.

\subsubsection{The Student's $t$ Statistic --- Testing Elements of $\Best$}
\label{sec:studentt}

We may also be interested in testing whether individual elements of
$\bv_{i}$ (rows of $\B$) are equal to zero. Each of the $k$ elements
of $\bv_{i}$ are coefficients defining a linear combination of PC
basis vectors $\Zt$ that construct each row of the feature matrix $\M$
linking physical parameters of rotating core collapse and principal
components (PCs).  Hypothesis tests on elements allow us to measure
how important individual PCs are to a given feature vector.

We use the complex form of the student's $t$ test statistic~\cite{Akaike1965,Brillinger1981}, given by
\begin{equation}
\label{eq:tau}
\tau = \frac{|\Best_{i,j}|^{2}}{\Cx_{ii} \hat{\Sig}_{Z_{jj}}}\,\,,
\end{equation}
where $\hat{\Sig}_{Z_{jj}}$ is the $j$th diagonal element of
$\hat{\Sig}_{Z}$.  For the real case,
see~\cite{Marden}.  Under $H_0$ ($\B_{i,j} = 0$), the distribution of
this test statistic is given by,
\begin{equation}
\label{eq:comptdist}
\frac{1}{2} \tau \sim \mathcal{F}_{2,2v} \,,
\end{equation}
where $2$ is the upper degrees of freedom parameter, and $2v$ is the
lower degrees of freedom parameter of the $\mathcal{F}$-distribution.
This test statistic can easily be used to produce circular confidence
intervals for each element of $\Best$ in the complex plane
(e.g., see Fig.~\ref{fig:pcsA1xbetaR}).

\subsubsection{Discussion of Test Statistics }
\label{sec:teststatdiscussion}

The complex forms of both the Hotelling's $T^2$ and the student's $t$
statistics are distributed according to the $\mathcal{F}$-distribution
(also known as the Fisher-Snedecor probability distribution,
see~\cite{JamesF}).  The factors of two in the degrees of freedom
parameters in Eqs.~\ref{eq:FtoP}~and~\ref{eq:comptdist} come from the
fact that our Fourier transformed waveforms are complex valued.  For a
derivation of Hotelling's $T^2$ statistic and student's $t$ statistic
in the real-valued case, see~\cite{Marden} and references
therein.  For the Hotelling's $T^2$ with complex data,
see~\cite{Giri1977}.

To compute $\eta$ in practice, the results of either
Eqs.~\ref{eq:Hotelling} or~\ref{eq:tau} are plugged into the left hand
side of either Eqs.~\ref{eq:FtoP} or~\ref{eq:comptdist}.  We label the
quantity obtained $\eta$.  Next, $\eta$ is transformed into a
$p$-value, which is more easily interpreted.  A $p$-value is the
probability, under the assumption that $H_{0}$ is true, of obtaining
an $\eta$ value as high as or higher than was computed.  For a more
detailed summary on the precise interpretation and computation of
$p$-values, see~\cite{PDG2012}. The $p$-value transform is defined as,
\begin{equation}
\label{eq:pvaluedef}
p\textrm{-value} = \int_{\eta}^{\infty} f(x; df_{upper}, df_{lower}) dx\,\,,
\end{equation}
where $f(x; df_{upper}, df_{lower})$ is the $\mathcal{F}$-distribution
function, $df_{upper}$ is the upper degrees of freedom, and
$df_{lower}$ is the lower degrees of freedom.  Keeping in mind that if
$H_0$ is true, $\eta$ values will be distributed according to the
probability distribution function $f(x; df_{upper}, df_{lower})$.
Therefore obtaining a small $p$-value indicates a lack of evidence for
$H_0$.  In this paper, we consider $p$-values at or below $0.01$
\emph{significant}, where \emph{significant} indicates that we reject
$H_0$ and favor $H_a$.

We note here that it is simple to alter our regression model for
waveforms that have not been Fourier transformed.  With real-valued
time domain waveforms, one would follow all the same procedures
described, but would drop the detector noise covariance matrix,
$\Sig_S$, and remove the factor of two from the degrees of freedom in
Eqs.~\ref{eq:FtoP} and~\ref{eq:comptdist}.  This is the only
alteration to the regression model and hypothesis testing method that
would need to be made in order to analyze, reconstruct, and predict
time domain waveforms.

\section{Statistical Analysis of the Abdikamalov~\emph{et al.} Waveform Catalog}
\label{section:analysis}

With relevant statistical modeling procedures accounted for, we move
on and present an analysis of the rotating core collapse GW signal
catalog of Abdikamalov~\emph{et al.}~(\cite{Abdikamalov2013} and
section~\ref{sec:AbCat}).  Before beginning an analysis, the set of
waveforms $\Y$ must be scaled to a common distance.  Throughout the
remainder of the paper, we scale all waveforms to the distance of
$10\,\mathrm{kpc}$ in each of our analyses.

Abdikamalov~\emph{et al.} \citep{Abdikamalov2013} studied how varying
rotational parameters (e.g., rotation parameter $\beta_{ic,b}$ of the
inner core at bounce and precollapse degree of differential rotation
$A$) affect the morphology of the emitted GWs.  Using a series of
design matrices, we shall gradually develop a multivariate regression
model of how changes in the rotational parameters correlate with waveform catalog morphology.  

Throughout the remainder of this paper, we use 7 PCs in our PC basis
$\Zt$ ($k = 7$) unless stated otherwise. This choice
is motivated by Logue \emph{et al.}~\cite{Logue2012}. Experiments with
more PCs show that the results remain essentially the same up to
$\sim$20 PCs, beyond which individual higher-order PCs contribute
little to the actual signal feature vectors and add degrees of freedom
that decrease the significance of results. We leave a more detailed
study of the sensitivity of our results to the number of employed PCs
to future work.

\subsection{Analyzing Differential Rotation}
\label{sec:DiffRotDev}

We begin our analysis of the Abdikamalov~\emph{et al.} waveform
catalog with comparisons of the waveforms grouped by their 5
differential rotation profiles in order to see how much they differ
from waveforms in the other groups on average.  This allows us to
measure the average difference between waveforms generated from
progenitors with different differential rotation setups.

The procedure to obtain these results, given in
Table~\ref{tab:DiffRotDummyEncoding}, is as follows: First, we
apply a dummy variable encoding on differential rotation and form four
different design matrices, each with a different reference group left
out (Section~\ref{sec:dummyvarencoding} details this step).  With the
first design matrix, we measure the significance of the difference
between the $A1$ and the $A2$ waveforms (denoted in
Tab.~\ref{tab:DiffRotDummyEncoding} as $A1 - A2$), the $A1$ and the
$A3$ waveforms, the $A1$ and the $A4$ waveforms, and the $A1$ and $A5$
waveforms.  In this design matrix, the $A1$ waveforms are the
reference group. The other three design matrices have $A2$, $A3$, and
$A4$ as their reference group, respectively, and account for all
remaining possible comparisons.

Under a dummy variable encoding of a parameter, the elements in each
row of $\Best$ are the PC coefficients that produce
the average difference between waveforms from progenitors with two differential
rotation profiles.  Hotelling's statistic (Eq.~\ref{eq:Hotelling})
tests all the elements of $\hat{\bv}_{i}$ simultaneously.  We list
both Hotelling's statistic, and the $p$-value derived from it.
Sometimes, we may find that two (or more) comparisons have highly
significant $p$-values that are numerically equivalent to zero.  In
this situation, the value of $T^2$ can be used to measure the
difference in significance between the two comparisons.

\begin{table}[!t]
\caption{ \small Results of pair-wise comparisons between waveforms
  with different differential rotation profiles.  An asterisk ($*$)
  marks results that are considered significant (large values of $T^2$
  producing $p$-values at or below 0.01 are considered ``significant'').
  The waveforms are all scaled to be at the common distance of 10~kpc.
  $Ai - Aj$ indicates that we are measuring the average difference
  between waveforms from cores with the $Ai$ differential rotation profile, and waveforms from cores with
  with the $Aj$ differential rotation profile. }
\label{tab:DiffRotDummyEncoding}
\begin{ruledtabular}
\begin{tabular}{l . l}
Comparison &
\multicolumn{1}{r}{Hotelling's $T^2$} &
\multicolumn{1}{c} {$p$-value} \\
	\hline   \rule{0 em}{1.2 em}%
	$A1 - A2$ & 26.63 & $4.4 \times 10^{-5}*$  \\
	$A1 - A3$ & 26.46 & $4.8 \times 10^{-5}*$  \\
	$A1 - A4$ & 23.78 & $2.1 \times 10^{-4}*$  \\
	$A1 - A5$ & 18.67 & $0.003*             $  \\	[0.5 em]
	$A2 - A3$ & 6.35 & $0.62               $  \\
	$A2 - A4$ & 16.22 & $0.01              *$  \\
	$A2 - A5$ & 17.01 & $0.008             *$  \\	[0.5 em]
	$A3 - A4$ & 5.58 & $0.73               $  \\
	$A3 - A5$ & 7.57 & $0.45               $  \\  [0.5 em]
	$A4 - A5$ & 0.98 & $0.999              $
\end{tabular}
\end{ruledtabular}
\end{table}

We find no evidence in Tab.~\ref{tab:DiffRotDummyEncoding} for a
significant difference between waveforms with differential rotation $A2$ and $A3$ ($A2 - A3$), $A3$ and $A4$ ($A3 - A4$), $A3$ and $A5$ ($A3 - A5$), as well as $A4$ and $A5$ ($A4 - A5$).  Differences are more significant for comparisons that involve waveforms from more differentially rotating progenitors.  Each comparison involving the $A1$ group is significant, and most of the comparisons involving $A2$ are as well.  This suggests that for a detected core collapse GW signal, it may be possible to determine either that its source was strongly differentially rotating (most similar to $A1$ or $A2$) or that its source had a more moderate degree of differential rotation (most similar to the $A3$, $A4$ and $A5$ parameterizations).

The significance of comparisons that involve $A1$ decreases as the
differential rotation of the comparison waveforms decreases.  This
does not necessarily suggest that $A1$ waveforms are more similar to
waveforms from more uniformly rotating progenitors than to those with
similar differential rotation profiles.  The $T^2$ value
(and therefore $p$-values transformed from it) is dependent not only on
the intrinsic difference between the waveforms in each of the groups
being compared, but also on the numbers of waveforms in each of the
groups.  There are 30 $A1$ waveforms, 22 $A2$ waveforms, 18 $A3$
waveforms, 12 $A4$ waveforms, and 10 $A5$ waveforms in the
Abdikamalov~\emph{et al.}  catalog.  As we remarked in
Sec.~\ref{sec:statistics}, the $\Cx_{ii}$ term in Hotelling's $T^2$ is
responsible for characterizing the relative scaling of the design
matrix columns.  There is more support for the significance of a
comparison if there is a large number of waveforms in each of the two
groups being compared.  The evidence for significance is driven down
when one (or both) of the groups in a comparison has a small number of
waveforms.  

To consider how influential different degrees of differential rotation
are individually, we examine how the GWs from each group compare to the overall catalog mean.  A deviation encoding allows us to measure how \emph{unique} a signature
in the waveforms produced with a given parameter value is, without
having to use a set of waveforms with another parameter value as a
reference.  This is accomplished with a deviation encoding of the
differential rotation parameter (see Sec.~\ref{section:encoding}).  In
Table~\ref{tab:DiffRotDevEncoding}, we list Hotelling's $T^2$ and the
corresponding $p$-value results of comparisons of the differential
rotation parameter groups with the catalog mean.  In
Tab.~\ref{tab:DiffRotDevEncoding}, the $\mu$ symbol denotes the
intercept term, the mean of all the catalog waveforms.

\begin{table}[!t]
\caption{ \small Testing the average difference between a set of
  waveforms partitioned by differential rotation profile and the mean
  of all catalog waveforms.  An asterisk ($*$) marks results that are
  considered significant (large values of $T^2$ producing $p$-values
  at or below 0.01 are considered ``significant'').  All waveforms are
  scaled to be at the common distance of 10~kpc.  Our results show
  that the $A1$ and to a lesser extent, the $A2$ waveforms are
  significantly different from the average of all catalog waveforms.}
\label{tab:DiffRotDevEncoding}
\begin{ruledtabular}
\begin{tabular}{l . l}
Comparison &
\multicolumn{1}{r}{Hotelling's $T^2$} &
\multicolumn{1}{c} {$p$-value} \\
	\hline   \rule{0 em}{1.2 em}%
	$A1 - \mu$ & 38.54 & $6.3 \times 10^{-8}*$    \\
	$A2 - \mu$ & 19.48 & 0.002*                 \\
	$A3 - \mu$ & 6.67 & 0.57                    \\
	$A4 - \mu$ & 7.67 & 0.44                    \\
	$A5 - \mu$ & 8.01 & 0.39                    \\

\end{tabular}
\end{ruledtabular}
\end{table}

The results in Tab.~\ref{tab:DiffRotDevEncoding} corroborate the
results in Tab.~\ref{tab:DiffRotDummyEncoding}.  We find that the $A1$ and $A2$ groups
indeed produce the most unique signature.  Waveforms from the $A1$
group are on average the most different from the mean of the catalog
waveforms (depicted in Fig.~\ref{fig:meanWF}).  This also supports the
conclusions about the impact of differential rotation drawn by
Abdikamalov~\emph{et al.}~\cite{Abdikamalov2013}.

\begin{figure}[t]
    \centerline{\includegraphics[width=8.6cm]{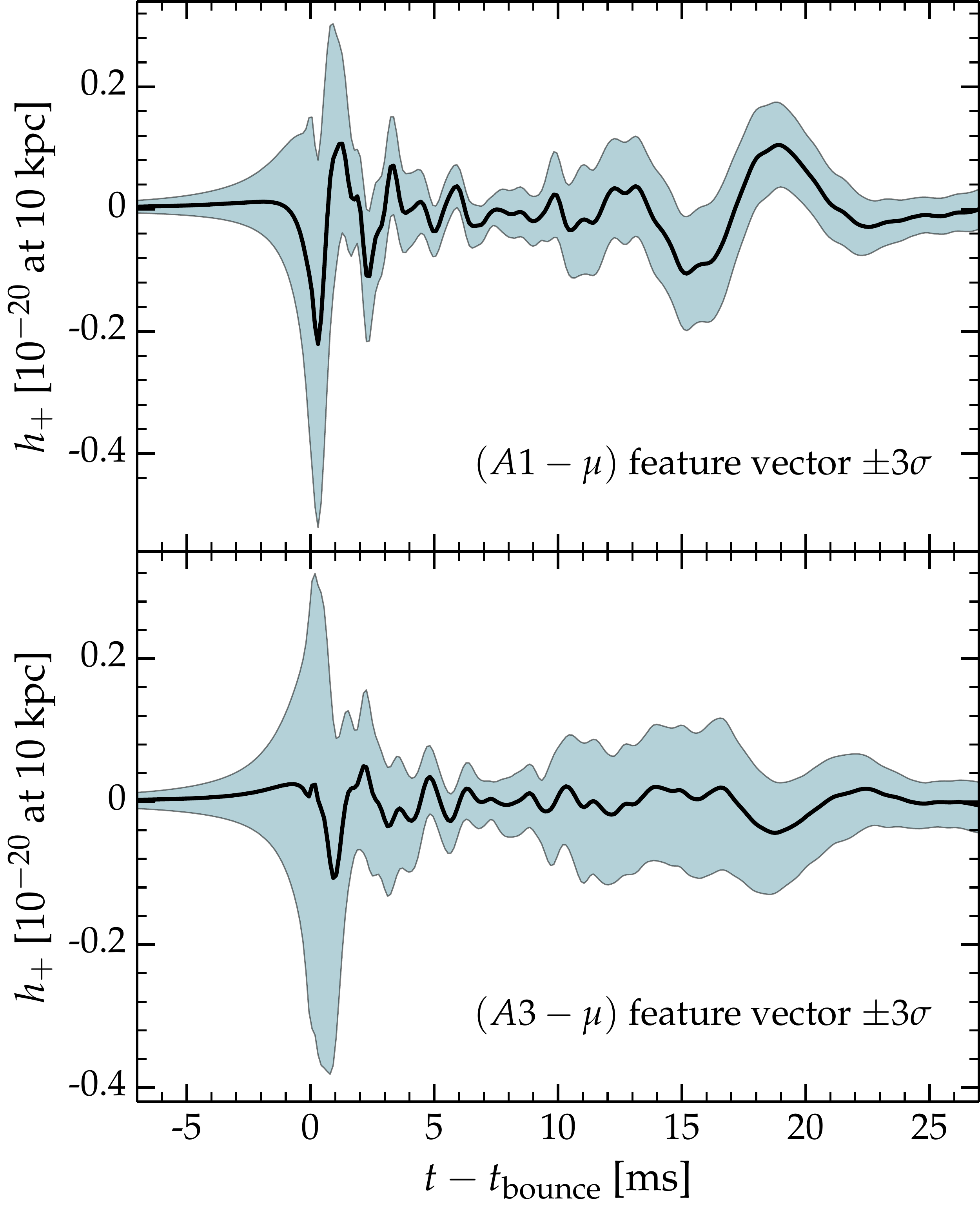}}
\caption{ \small Two time domain feature vectors shown with a $3
  \sigma$ confidence region produced using the deviation encoded
  design matrix used in Tab.~\ref{tab:DiffRotDevEncoding}.  The top
  panel shows the $A1 - \mu$ feature vector.  The large
    amplitudes between about $10$ and $20$ milliseconds in the $A1$
    feature vector suggests that the $A1$ waveforms differ
    significantly from the catalog mean in that phase.  The bottom
    panel shows the $(A3 - \mu)$ feature vector.  The wider confidence
    region indicates the lack of a robust feature vector that can be
    used to characterize the difference between the $A3$ waveforms
    from the catalog mean.  To produce these feature vectors, the
    waveforms in the catalog were originally scaled to a distance of
    10 kpc. }
\label{fig:tdfeatures}
\end{figure}

In order to visualize the results of
Tab.~\ref{tab:DiffRotDevEncoding}, we estimate the uncertainty of
$\hat{\M}$ in the time domain using the estimated standard deviations
of the elements of $\Best$, given by $\Cx_{ii} \hat{\Sig}_{Z_{jj}}$.
For the comparisons listed in any of our tables that have lower
$p$-values, we can expect to see smaller estimated errors in their
corresponding estimated feature vectors.  The top panel of
Fig.~\ref{fig:tdfeatures} shows the feature vector that corresponds to
the $A1 - \mu$ column of a design matrix comprised of a deviation
encoding on differential rotation.  When testing the row of $\Best$
that produces this feature vector, we obtain a $p$-value of $6.3
\times 10^{-8}$ (the first row of Tab.~\ref{tab:DiffRotDevEncoding}).
The bottom panel of Fig.~\ref{fig:tdfeatures} is the feature vector
that represents $A3 - \mu$, for which we obtain a $p$-value of 0.57.
The $A1 - \mu$ feature vector is the most significant in
Tab.~\ref{tab:DiffRotDevEncoding}, and the $A3 - \mu$ feature vector
is the least significant.  Both time domain feature vectors are
plotted with $3 \sigma$ error regions.  As the $p$-value results
suggest, the $A1 - \mu$ time domain feature vector has both a larger
amplitude and a narrower error region.  

\subsection{The Influence of Total Rotation}
\label{sec:totalrotresults}

Abdikamalov~\emph{et al.}~\citep{Abdikamalov2013} observed that the
morphology of the waveforms in their catalog is highly dependent on
the ratio of rotational kinetic energy to gravitational energy of the
inner core at bounce, $\beta_{ic,b}$, where the subscript $_{ic,b}$
stands for ``inner core, at bounce''.  This parameter is a good
measure of the progenitor core's \emph{total
  rotation}~\cite{Abdikamalov2013}, and continuously varies from
$\beta_{ic,b} = 0.0016$ to $\beta_{ic,b} = 0.206$ throughout the
Abdikamalov~\emph{et al.} catalog.  In this section, we examine
results using design matrices parameterized by total rotation.  We bin
$\beta_{ic,b}$ into three groups, corresponding to slow, moderate and
rapid rotation.  We use the labels S, M, R to denote this:
\begin{itemize}

	\item $\beta S = [0.0016, 0.0404]$, 30 waveforms;

	\item $\beta M = [0.0414, 0.1096]$, 31 waveforms;

	\item $\beta R = [0.115, 0.206]$, 31 waveforms.

\end{itemize}
We choose these ranges based on Fig.~10 of Abdikamalov~\emph{et
  al.}~\citep{Abdikamalov2013}.  These ranges are approximately ranges
over which $\beta_{ic,b}$ produces qualitatively similar behavior in
three of the primary waveform peaks~\citep{Abdikamalov2013}.

We begin an analysis of total rotation by using a dummy variable
encoding on our three total rotation ranges.  The results of this
encoding are shown in Table~\ref{tab:TotalDummyEncoding}.  The results
in this table show that total rotation is much more
influential on GW morphology than differential rotation.  The values
of $T^2$ (and their $p$-values) show a dramatic increase in
significance compared to the results in
Tables~\ref{tab:DiffRotDummyEncoding}
and~\ref{tab:DiffRotDevEncoding}.  This means that differences in
waveform morphology are much more pronounced when partitioning
waveforms by $\beta_{ic,b}$.  The $p$-values obtained for every
comparison are equal to zero, to machine precision, and the values of
Hotelling's $T^{2}$ are exceptionally large.

These results suggest that parameter estimation methods should be able
to accurately measure the total rotation from a rotating core collapse
GW signal detected by Advanced LIGO.  This is in agreement with
Abdikamalov~\emph{et al.}~\cite{Abdikamalov2013}, who use a match
filtering parameter estimation approach~\cite{Finn1992} to measure
$\beta_{ic,b}$ to within $\sim 30\%$ of its true value. They also show
that $\beta_{ic,b}$ can be directly related to the total angular
momentum of the inner core at bounce.  Thus the ability to measure
$\beta_{ic,b}$ provides a straightforward way to determine the angular
momentum content in the core of a collapsing star.

\begin{table}[t]
\caption{ \small Results of comparisons between waveforms partitioned
  into three groups based on $\beta_{ic,b}$, a parameter
  expressing the total rotation of the inner core at bounce.  While
  all comparisons marked with an asterisk ($*$) are significant
  ($p$-value $\leq 0.01$), a larger value of $T^{2}$ can be used to
  determine how different from each other waveforms from different
  groups are, since all comparisons produced $p$-values numerically
  equivalent to zero.  All waveforms are scaled to a distance of 10
  kpc.  $\beta i$ indicates one of three ranges of $\beta_{ic,b}$ (see text for details).  $\beta i - \beta j$ indicates that we are measuring the
  average difference between the sets of waveforms from progenitors with the $\beta i$ and
  the sets of waveforms from progenitors with the $\beta j$ total rotation.  }
\label{tab:TotalDummyEncoding}
\begin{ruledtabular}
\begin{tabular}{lcc}
	Comparison           &
	Hotelling's $T^{2}$              &
	$p$-value            \\
	\hline   \rule{0 em}{1.2 em}%
	$\beta S - \beta M$ & 132.7 & $0.0*$   \\
	$\beta S - \beta R$ & 311.7 & $0.0*$   \\
	$\beta M - \beta R$ & 205.0 & $0.0*$   \\
\end{tabular}
\end{ruledtabular}
\end{table}
\begin{table}[!t]
\caption{ \small Results of comparisons between waveforms grouped by different ranges of $\beta_{ic,b}$ and values of $A$, and the catalog mean.  Both parameters
  were simultaneously encoded in the design matrix.  The waveform
  catalog is originally scaled to a distance of 10 kpc.  $A i - \mu$
  or $\beta i - \mu$ indicates that we are measuring the average
  difference between that set of waveforms and the average of all
  catalog waveforms.  An asterisk ($*$) marks results that are
  considered significant (large values of $T^2$ producing $p$-values
  at or below 0.01 are considered ``significant'').}
\label{tab:TotalDevEncoding}
\begin{ruledtabular}
\begin{tabular}{l . l}
Comparison &
\multicolumn{1}{r}{Hotelling's $T^2$} &
\multicolumn{1}{c} {$p$-value} \\
	\hline   \rule{0 em}{1.2 em}%
	$A1 - \mu$ & 49.7 & $2.0 \times 10^{-10}*$   \\
	$A2 - \mu$ & 18.1 & $4.4 \times 10^{-3}*$    \\
	$A3 - \mu$ & 9.2 & 0.27                    \\
	$A4 - \mu$ & 8.5 & 0.34                  \\
	$A5 - \mu$ & 6.0 & 0.67                    \\  [0.5 em]
	$\beta S - \mu$ & 260.4 & $0.0*$   \\
	$\beta M - \mu$ & 117.8 & $0.0*$   \\
	$\beta R - \mu$ & 309.6 & $0.0*$
\end{tabular}
\end{ruledtabular}
\end{table}

Next, we test solutions from design matrices that are a concatenation
of a deviation encoding on the three ranges of $\beta_{ic,b}$, and a
deviation encoding on the five levels of differential rotation ($A1$
through $A5$).  For more details on this type of procedure, see
Section~\ref{sec:multipleparam+interactions}. This scheme improves our
inferences on both the differential and total rotation parameters
because it produces a solution where the effects of the two types of
parameters on GW morphology are separated.  By using a concatenated
design matrix, feature vectors contain \emph{only} morphology relevant
to either $A$ or $\beta_{ic,b}$.

In Table~\ref{tab:TotalDevEncoding}, we list results from this
encoding.  As the strength of differential rotation decreases, the
significance decreases (the $p$-values become larger).  These results
are more trustworthy than those given in
Tab.~\ref{tab:DiffRotDevEncoding}, because the effects on the
waveforms due to $\beta_{ic,b}$, which are found to be much stronger
than those due to differential rotation, have been factored out.

\subsection{Interactions Between Differential and Total Rotation}
\label{sec:interactions}

Abdikamalov~\emph{et al.}~\cite{Abdikamalov2013} find evidence for
important inter-dependencies between differential rotation and total
rotation.  For slowly rotating progenitors leading to
  $\beta_{ic,b} \lesssim 0.04$ to $0.08$, the waveforms are
  essentially independent of differential rotation. Only at higher
  values of $\beta_{ic,b}$ is differential rotation influential on
  the GW signal shape.

In order to examine the dependencies between total and differential
rotation, we can encode \emph{two-way interactions} between the
differential and total rotation parameters.  A two-way interaction
means waveforms are grouped by two parameters, allowing their joint
effect on waveform morphology to be recovered (see
Sec.~\ref{sec:multipleparam+interactions} for a detailed explanation).
For instance, we may consider waveforms with $\beta_{ic,b} \lesssim
0.05$ and the $A1$ differential rotation as a single group, and then
test whether these waveforms have a distinct morphology.

Results from
Tables~\ref{tab:DiffRotDummyEncoding},~\ref{tab:DiffRotDevEncoding}~and~\ref{tab:TotalDevEncoding}
suggest that waveforms with $A3$, $A4$ and $A5$ differential rotation
profile can be grouped together, due to the lack of evidence for
significant differences between these groups.  In order to reflect
this new grouping, we alter the differential rotation parameter
labeling, using the letter 'U' to reflect that these waveforms are
from uniformly to moderately differentially rotating progenitors:
\begin{itemize}

	\item  $A1$ = $A1$, 30 waveforms;

	\item  $A2$ = $A2$, 22 waveforms;

	\item  $AU$ = $A3$, $A4$ and $A5$, 40 waveforms.

\end{itemize}
\begin{figure}[t]
    \centerline{\includegraphics[width=8.6cm]{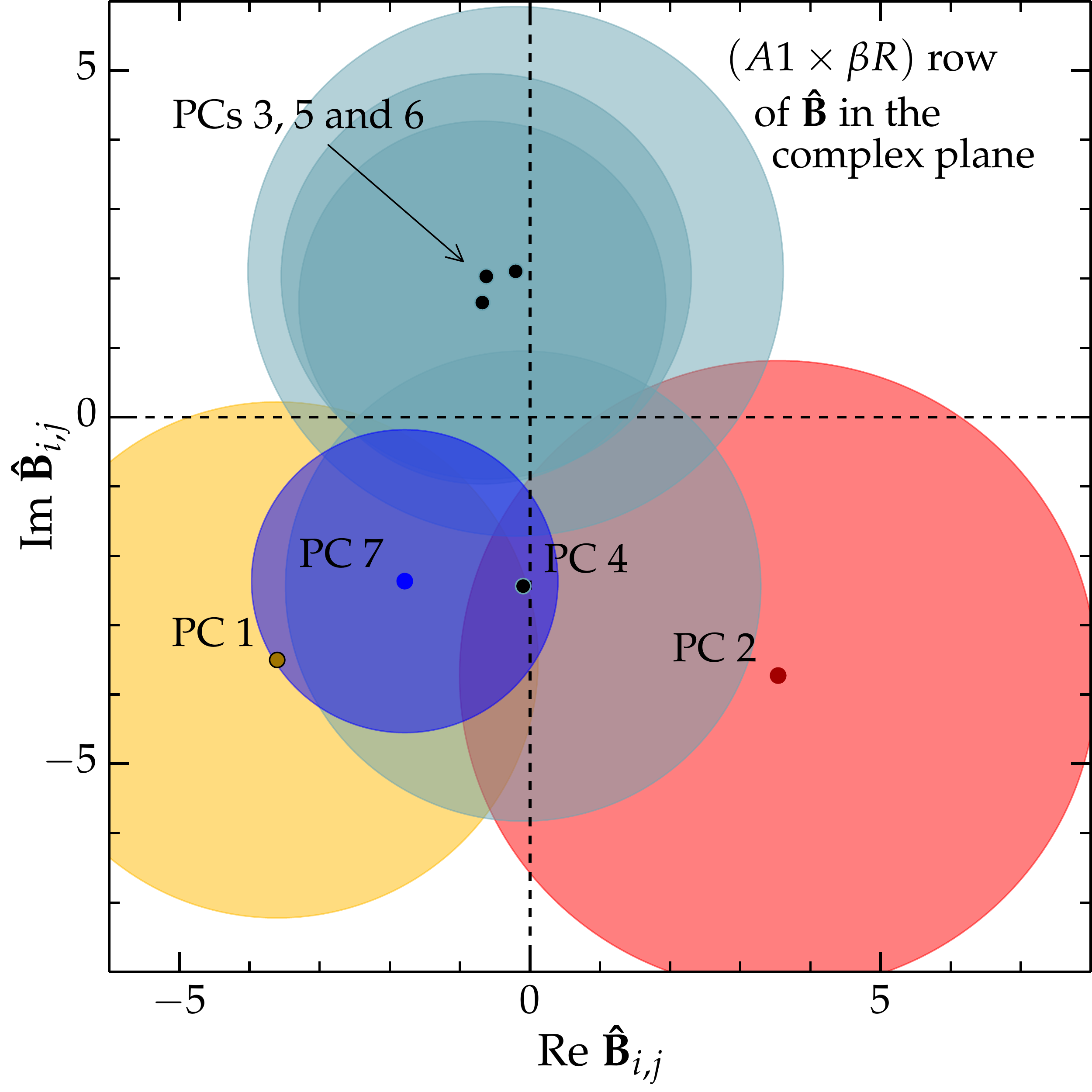}}
\caption{\label{fig:pcsA1xbetaR} \small 95\% Confidence circles in the
  complex plane for the $i$th row of $\Best$, which contains the PC
  coefficients of the $(A1 \times \beta R)$ interaction feature
  vector.  The column of the design matrix $(A1 \times \beta R)$ was
  encoded into determines the value of $i$. The $(A1 \times \beta R)$
  feature vector describes waveforms that are both highly
  differentially rotating ($A1$) and have a rapid total rotation
  ($\beta R$).  The PC coefficients of row $\Best_{i}$ are marked in
  black.  The $j = 3,4,5,6$ PC coefficients overlap the origin and
  their 95\% confidence circles are shaded in subdued colors.  From
  this plot, we can see that the $(A1 \times \beta R)$ feature vector
  is primarily determined by the $j = 1,2$ and $7$ PCs, whose
  confidence circles do not overlap zero. }
\end{figure}
\begin{table}[t]
\caption{ \small Results of comparisons of two-way interactions
  between waveforms grouped into three differential rotation ($A$)
  categories, and into three ranges of total rotation
  ($\beta_{ic,b}$).  The only set of interactions that are found to be
  not significant ($p$-value~$\geq~0.01$) are those involving
  waveforms with the $A2$ differential rotation profile.  All catalog
  waveforms were scaled to a distance of 10 kpc.  An asterisk ($*$)
  marks results that are considered significant (large values of $T^2$
  producing $p$-values at or below 0.01 are considered
  ``significant''). }
\label{tab:Interactions}
\begin{ruledtabular}
\begin{tabular}{l . l}
Comparison &
\multicolumn{1}{r}{Hotelling's $T^2$} &
\multicolumn{1}{c} {$p$-value} \\
	\hline   \rule{0 em}{1.2 em}%
	$A1 - \mu$ & 64.9 & $1.4 \times 10^{-13}*$   \\
	$A2 - \mu$ & 21.57 & $7.5 \times 10^{-4}*$   \\
	$AU - \mu$ & 39.88 & $3.9 \times 10^{-8}*$   \\ [0.3 em]

	$\beta S - \mu$ & 353.52 & $0.0*$   \\
	$\beta M - \mu$ & 157.53 & $0.0*$   \\
	$\beta R - \mu$ & 561.72 & $0.0*$   \\ [0.6 em]

	$A1 \times \beta S$ & 36.40 & $2.5 \times 10^{-7}*$   \\
	$A1 \times \beta M$ & 36.10 & $2.9 \times 10^{-7}*$   \\
	$A1 \times \beta R$ & 71.94 & $5.6 \times 10^{-15}*$   \\  [0.3 em]

	$A2 \times \beta S$ & 6.23  & $0.64$   \\
	$A2 \times \beta M$ & 7.79  & $0.42$   \\
	$A2 \times \beta R$ & 10.72 & $0.15$   \\  [0.3 em]

	$AU \times \beta S$ & 32.40 & $2.2 \times 10^{-6}*$   \\
	$AU \times \beta M$ & 31.63 & $3.3 \times 10^{-6}*$   \\
	$AU \times \beta R$ & 44.92 & $2.8 \times 10^{-9}*$   \\

\end{tabular}
\end{ruledtabular}
\end{table}

Our partitioning of the physical parameters into three different
differential rotation ranges and three total rotation ranges leads to
nine different two-way interactions to test, in addition to six tests
of the deviation encoding on $A$ and $\beta_{ic,b}$.  The results are
given in Table~\ref{tab:Interactions}.  We find that all $p$-values
are lower than $0.01$, except those for interactions involving the
$A2$ waveforms. 

Therefore, there is no evidence for a strong inter-dependence of $A2$
waveforms on $\beta_{ic,b}$ --- the three features for the $A2$ with
$\beta S$, $\beta M$ and $\beta R$ waveforms are not significant.  The
changes in the $A2$ waveforms due to $\beta_{ic,b}$ are better
explained by the $\beta S - \mu$, $\beta M - \mu$ and $\beta R - \mu$
features.  This is not the case for the other differential rotation
levels, whose waveforms as a whole exhibit varying, but generally
strong degrees of inter-dependence with $\beta_{ic,b}$.

Since rotating core collapse is a highly non-linear process, it is not
surprising to find strong inter-dependencies between these two
parameters.  To highlight the connection of our work to the PC-based
methods of Heng~\cite{Heng2009} and R\"over~\emph{et
  al.}~\cite{Rover2009}, we use student's $t$ statistic to examine the
importance of individual principal components (PCs) in one of the
interaction terms.  The two-way interaction between $A1$ and $\beta
R$, labeled $A1 \times \beta R$ in Table~\ref{tab:Interactions},
resulted in the lowest $p$-value of the interactions tested, $5.6
\times 10^{-15}$.  Abdikamalov~\emph{et al.}~\cite{Abdikamalov2013}
also find that the distribution of angular momentum (differential
rotation) is most relevant to the GW signal for very rapidly rotating
cores (high $\beta_{ic,b}$).

In order to visualize the solutions (rows of $\Best$) obtained by our
regression approach, we plot confidence intervals around the PC
coefficients used to reconstruct waveforms in the $A1 \times \beta R$
waveform group in Fig.~\ref{fig:pcsA1xbetaR}.  From
  Fig.~\ref{fig:pcsA1xbetaR}, we find that PCs 1, 2, and 7 are
  primarily responsible for uniquely characterizing the set of
  waveforms that were generated from strongly differentially rotating
  progenitors with rapidly rotating cores.

\subsection{Ability of the Model to Reconstruct Waveforms}
\label{sec:polyresults}

In this section, we again use a deviation encoding to model the
differential rotation parameter, and for $\beta_{ic,b}$, transition to
the use of a polynomial encoding.  For the time being, we neglect two-way
interaction terms between polynomials of $\beta_{ic,b}$ and
differential rotation.  The polynomial encoding of $\beta_{ic,b}$ is
useful for associating trends in GW morphology with changing values of
$\beta_{ic,b}$.  While results can be more difficult to interpret in
an analysis due to the multivariate nature of the waveforms, polynomial
terms can still provide insight into waveform morphology.

Encoding the continuous valued $\beta_{ic,b}$ parameter with
polynomials also avoids the need to specify bin ranges.  For
continuous parameters, it is generally difficult to choose the number
of bins and the range each bin covers.

Higher order polynomials in the design matrix are also a good way to
obtain accurate reconstructions of catalog waveforms.  We build a fifth
order polynomial model for the $\beta_{ic,b}$ parameter to see how
well our model can fit the catalog.  If there are $n$ data points on
some two dimensional scatter plot, an $n$th order polynomial is
required to exactly fit the data points~\cite{Tibshirani}.  This logic
applies in the multivariate case as well.  With $n$ waveforms, an
$n$th order polynomial can provide a perfect fit.  We use a 5th
order polynomial of $\beta_{ic,b}$ that is flexible enough to fit
shapes similar to those in Fig. 10 of Abdikamalov~\emph{et al.}, but also has a low enough order to avoid oscillations between interpolated points associated with high-order polynomials (Runge's phenomenon).

\begin{figure}[t!]
    \centerline{\includegraphics[width=8.6cm]{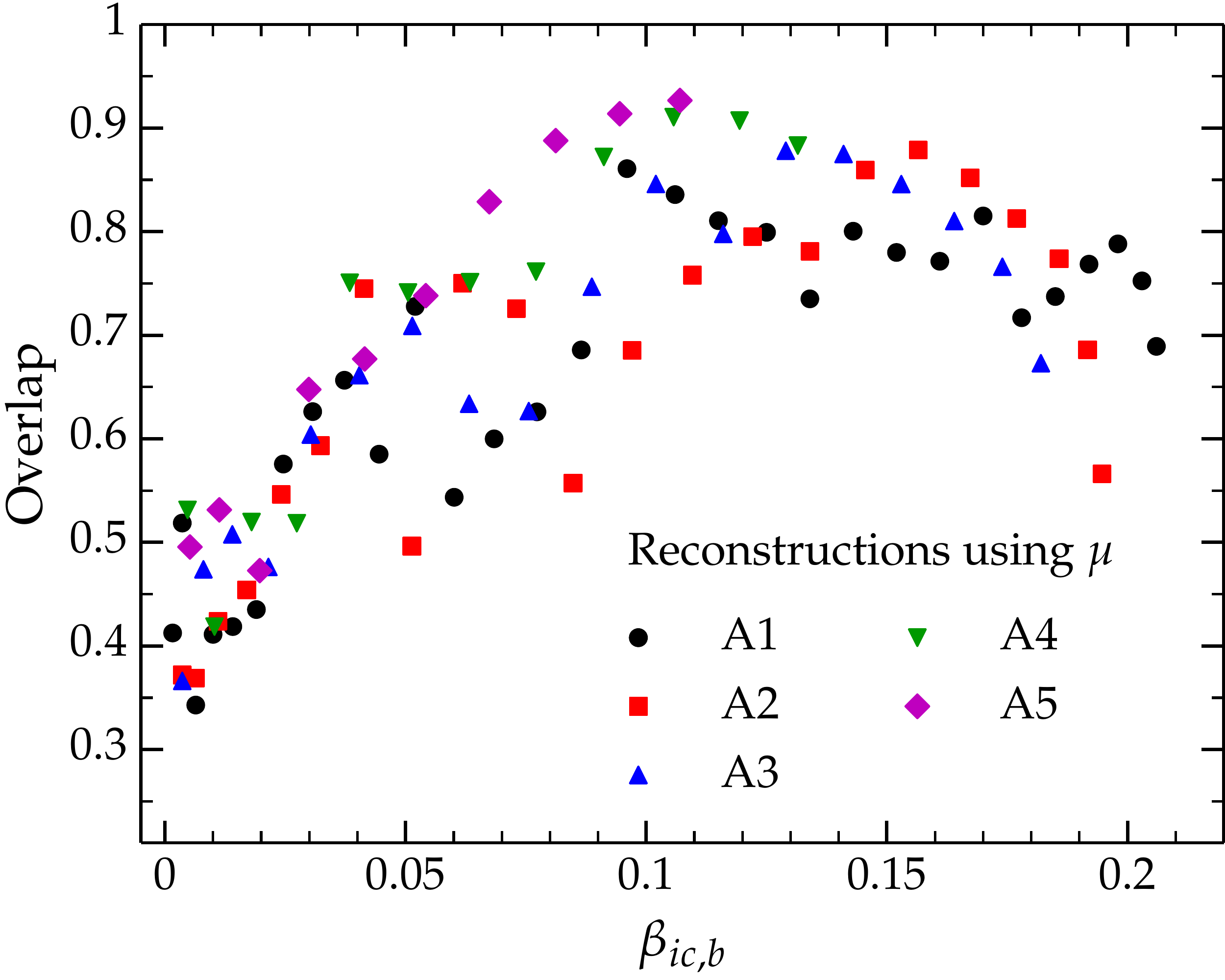}}
\caption{\label{fig:meanmatch} \small Overlap as a function of
  $\beta_{ic,b}$ for the Abdikamalov~\emph{et
    al.}~\cite{Abdikamalov2013} waveforms using only the catalog mean
  (a design matrix with only a column of ones, denoted $\mu$) to
  reconstruct the 92 primary waveforms.  The differential rotation is
  represented by various marker types.  }
\end{figure}
\begin{figure}[t!]
    \centerline{\includegraphics[width=8.6cm]{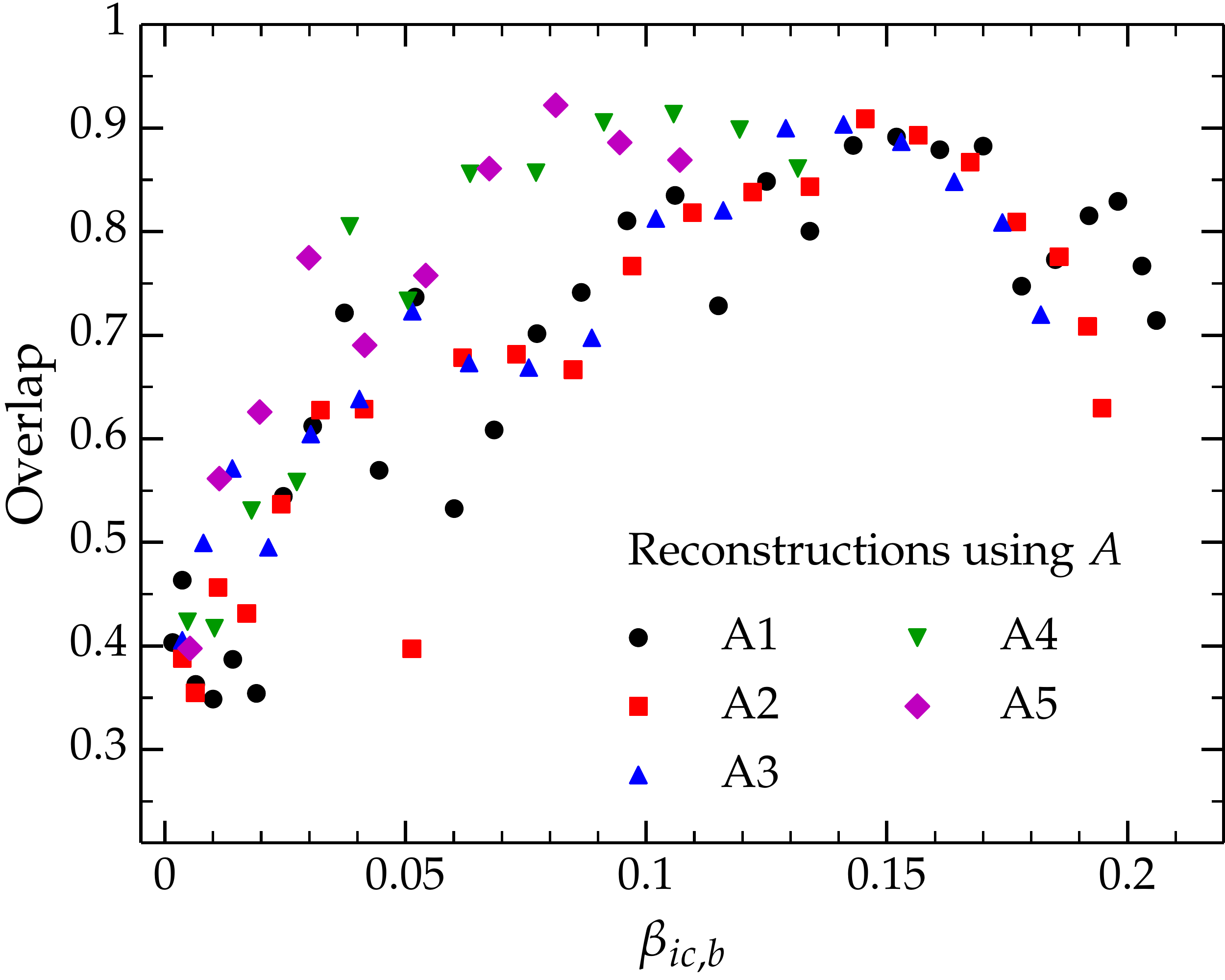}}
\caption{\label{fig:adevia} \small Overlap as a function of
  $\beta_{ic,b}$ for the Abdikamalov~\emph{et
    al.}~\cite{Abdikamalov2013} waveforms with a deviation encoding on
  differential rotation ($A$) to reconstruct the 92 primary waveforms.
  Each waveform is reconstructed by the mean waveform and a feature
  vector associated with a particular differential rotation profile.
  Slight improvements in overlap from Fig.~\ref{fig:meanmatch} are
  noticeable.  }
\end{figure}

After forming a design matrix $\X$ with a deviation encoding of
differential rotation and a polynomial encoding on $\beta_{ic,b}$, we
solve for $\Best$ and use it to reconstruct all catalog waveforms.  We
then find the set of reconstructed waveforms, denoted $\Y^{R}$,
by simply plugging $\Best$ into
\begin{equation}
\Y^{R} = \X \Best \Zt \,,
\end{equation}
along with the appropriate design matrix $\X$ and PC basis $\Zt$.

The criterion we use to determine the accuracy of reconstructions (or
predictions) is the detector noise weighted overlap.  An overlap of
one means two waveforms are identical, while an overlap of zero
indicates that they are orthogonal.  To compute the overlap, we first
define the detector noise weighted inner product,
\begin{equation}
< g , h > \hspace{2mm} = \hspace{2mm} 2 \int_{0}^{\infty}  df \frac{\tilde{g}(f)\tilde{h}^{*}(f) + \tilde{g}^{*}(f)\tilde{h}(f)}{S_n(f)} \,,
\end{equation}
where $\tilde{h}_{k}(f)$, $\tilde{g}_{k}(f)$ are the Fourier
transforms of $h(t)$ and $g(t)$, two signals we are interested in
comparing.  The~$*$~denotes complex conjugation, and $S_n(f)$ is the
known detector noise power spectral density.  The overlap,
$\mathcal{O}_{i}$, of the $i$th waveform, $\yv_{i}$ with its
reconstruction, $\yv_{i}^{R}$, is defined as
\begin{equation}
\label{match}
\mathcal{O}_{i} \equiv \frac{ <\yv_{i}^{R}, \yv_{i}>  }{ \sqrt{
<\yv_{i}^{R}, \yv_{i}^{R}>
<\yv_{i}, \yv_{i}>  } } \,\,,
\end{equation}
which equals one if the two waveforms are entirely in phase, and is
zero when they are completely out of phase, where we are keeping the
waveforms perfectly aligned throughout.

\subsubsection{Reconstructions using the catalog mean and differential rotation}

We plot four different sets of reconstructions.  First, we use only
the intercept term $\mu$ (the first column of $\X$ in all of our
encoding schemes).  It can be shown that with only a column of ones in
$\X$, $\X \Best \Zt$ is equal to the mean waveform of the catalog,
which we denote $\bar{\yv}$.  This mean waveform (in the time domain)
is plotted in black in Fig.~\ref{fig:meanWF}, and is alternatively found
by taking the sum over all columns of $\Y$ and then dividing by the
total number of rows,
\begin{equation}
\bar{\yv} = \frac{1}{n}\sum_{j = 1}^{n} \Y_{j} \,.
\end{equation}
In this case, $\bar{\yv} = \yv_{i}^{R}$ for all $n$ catalog waveforms.
The overlap value for each waveform is plotted as a function of
$\beta_{ic,b}$ in Fig.~\ref{fig:meanmatch}.  Using $\bar{\yv}$ to
reconstruct,  48 out of 92 waveforms ($\sim 52\%$) have an overlap
greater than or equal to 0.7, indicating that many of the catalog
waveforms share a similar general form.  We also observe that
waveforms with $\beta_{ic,b}~\lesssim~0.1$ are much more difficult to
reconstruct, most likely because they contain stochastic signal
features from convection.  To a lesser extent, waveforms from rapidly
rotating progenitors, $\beta_{ic,b}~\gtrsim 0.15~$, are also more
unlike $\bar{\yv}$.  There appears to be no clear and visible
indication of a dependence of overlap on differential rotation, whose
values are denoted in Fig.~\ref{fig:meanmatch} by the colored symbols.

Next, in Fig.~\ref{fig:adevia}, we solve for $\Best$ using the
intercept ($\mu$) and the four deviation encoded columns for
differential rotation.  There is a small but noticeable improvement in
the reconstruction errors.  In this case, 53 waveforms out of 92 have
an overlap greater than 0.7 ($\sim$ 58\%).  Again, there appears to be
more difficulty in reconstructing waveforms from more slowly or more
rapidly rotating progenitors, but no obvious dependence on
differential rotation.

\subsubsection{Improving Reconstructions by Incorporating $\beta_{ic,b}$ and Two-way Interactions}

\begin{figure}[t!]
    \centerline{\includegraphics[width=8.6cm]{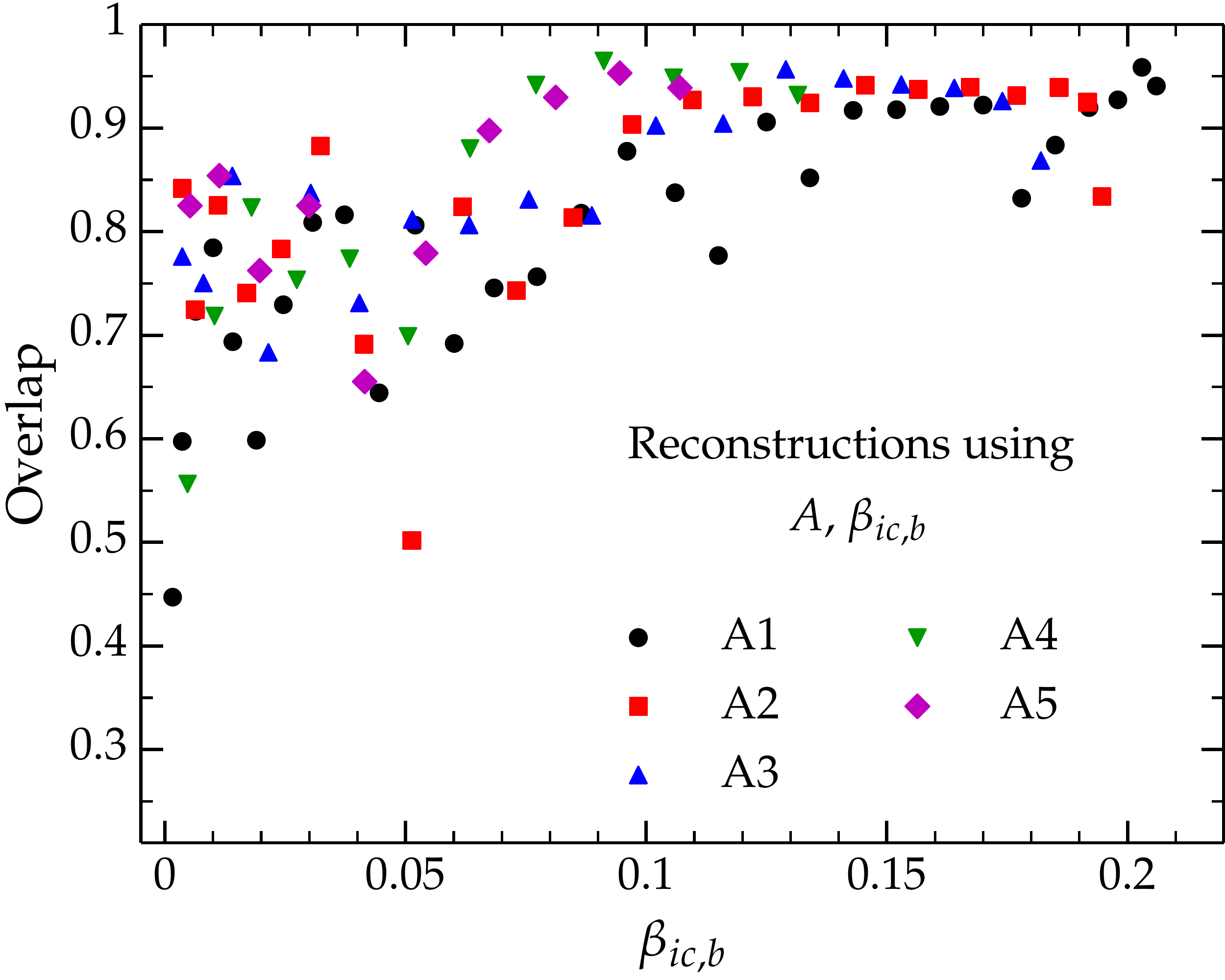}}
\caption{\label{fig:abetamatch} \small Overlap as a function of
  $\beta_{ic,b}$ for the 92 Abdikamalov~\emph{et
    al.}~\cite{Abdikamalov2013} waveforms.  A deviation encoding of
  $A$, as well as a 5th order polynomial function of $\beta_{ic,b}$,
  is encoded and fit.  Including the $\beta_{ic,b}$ parameter in the
  design matrix produces a large increase in the overlaps over the
  encoding used in Fig.~\ref{fig:adevia}.  }
\end{figure}

We include a $5$th order polynomial on $\beta_{ic,b}$ in the design
matrix, in addition to a deviation encoding of differential rotation
(both encodings necessitate the inclusion of a column of ones ($\mu$)
in the design matrix).  This encoding provides a dramatic increase in
the overlap between the waveforms and their reconstructions, as shown
in Fig.~\ref{fig:abetamatch}.  The reconstructions are excellent for
waveforms with $\beta_{ic,b}~\gtrsim~0.1$.  In total, 83 of the
waveforms now have an overlap greater than or equal to 0.7 ($\sim$
90\%).  This improvement corroborates our findings using $p$-values
about the strength of the correlation between GW morphology and total
rotation. Interestingly, there is a kink in the overlaps near
$\beta_{ic,b}~\sim~0.05$, indicating a point in the progenitor
parameter space whose waveforms are particularly difficult to
reconstruct.  We note from Fig. 10 in Abdikamalov~\emph{et al.} that
when $\beta_{ic,b} \approx 0.05$, the amplitude of the waveforms'
largest peak (the bounce peak, denoted $h_{1,neg}$) begins to change
as $A$ varies.  Both of our results indicate that $\beta_{ic,b}
\approx 0.05$ is a particularly volatile point in the parameter space
of rotating core collapse.

\begin{figure}[t!]
    \centerline{\includegraphics[width=8.6cm]{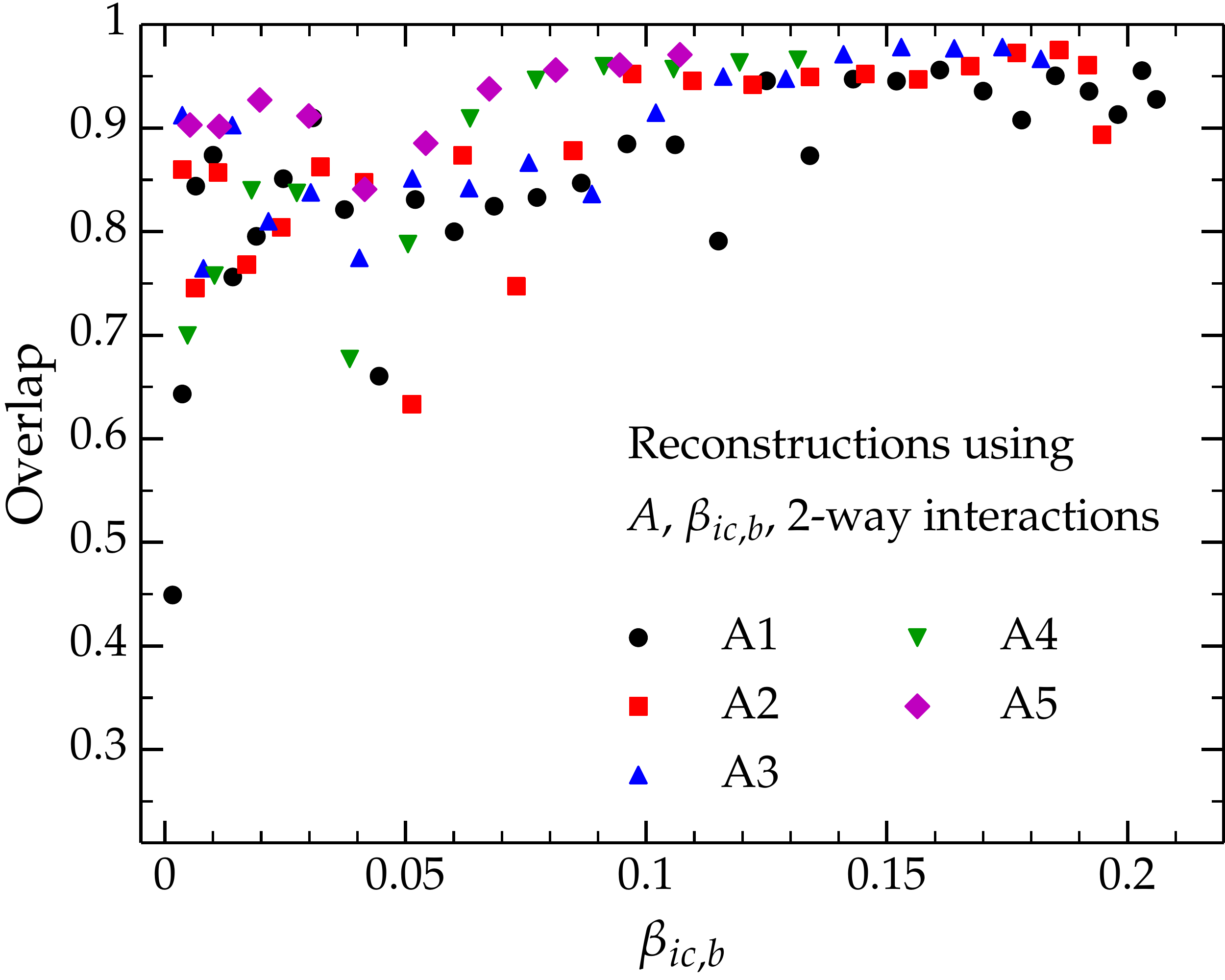}}
\caption{\label{fig:a7polyintera} \small Overlap as a function of
  $\beta_{ic,b}$ for the 92 Abdikamalov~\emph{et al.} waveforms.  This
  time, we use a deviation encoding of $A$, a 5th order polynomial
  function of the $\beta_{ic,b}$, as well as interactions between each
  of the 5 polynomial terms and the $A$ parameter.  This encoding
  produces the most accurate reconstructions of the catalog waveforms
  for the encodings we examine.  }
\end{figure}

While including a polynomial encoding of $\beta_{ic,b}$ improves the
overlap, waveforms from slowly rotating progenitors are still less
accurately reconstructed.  This is suggestive of two things.  First,
slowly spinning models emit GW signals with stronger stochastic
effects due to prompt postbounce convection~\cite{Abdikamalov2013,
  Dimmelmeier2008}.  This effect is problematic for our statistical
analysis due to the form of the Hotelling's $T^{2}$ and student's $t$
test statistics.  Both of these statistics are weighted by the
residual covariance matrix, $\Sig_{R}$, which is solved for using the
entire waveform catalog.  This procedure implicitly assumes that the
residuals of waveforms comprising the entire parameter space have the
same covariance structure.  We leave a detailed analysis of the
covariance structure of the residuals for further work.  Second, a
$5$th order polynomial model may provide an inadequate description for
waveforms from slowly rotating progenitors.  A higher-order
polynomial, or a different type of basis function may be required to
accurately capture the variation in the waveforms from more slowly
rotating progenitors.

Next, we build a design matrix that includes interactions between $A$
and $\beta_{ic,b}$.  This design matrix has one column in $\X$ for
$\mu$, four columns for a deviation encoding of $A$, five columns for
the $5$th order polynomial function of $\beta_{ic,b}$, and 20
interaction columns between each term in the $\beta_{ic,b}$ encoding
and each term in the $A$ encoding.  Including interactions results in
large overlaps for nearly all the waveforms in the
Abdikamalov~\emph{et al.}~\cite{Abdikamalov2013} waveform
catalog. This is shown in Fig.~\ref{fig:a7polyintera}.  Of the 92
primary waveforms, 88 have an overlap greater than or equal to 0.7
($\sim$ 96\%).  Most of the waveforms ($\sim$ 57\%) even have an
overlap $\gtrsim$ 0.9.  Again, most of these are from moderate to
rapid rotators with $\beta_{ic,b} \gtrsim 0.06 - 0.08$.  We also note
that the kink at $\beta_{ic,b} \sim 0.05$ in
Fig.~\ref{fig:a7polyintera} has become somewhat more pronounced.

\subsection{Predicting Injection Waveforms}
\label{sec:oos}

There is always the chance that our statistical model will be unable
to generalize to waveforms with parameterizations not specifically
encoded in the design matrix.  Alongside their primary catalog of 92
waveforms, Abdikamalov~\emph{et al.}~\cite{Abdikamalov2013} also
produced a set of 43 waveforms to be used as~\emph{injections}.  They
were used to test the ability of matched filtering and Bayesian model
selection methods to measure the physical parameters of GWs injected
into simulated detector noise.

To evaluate the ability of our regression model to predict waveforms,
we take the subset of 31 injection waveforms that does not include
waveforms computed with equations of state and electron capture
prescriptions that differ from those of the original catalog.  We do
this to simplify our analysis and will address dependence on equation
of state and electron capture microphysics in future work.
\begin{figure}[t!]
    \centerline{\includegraphics[width=8.6cm]{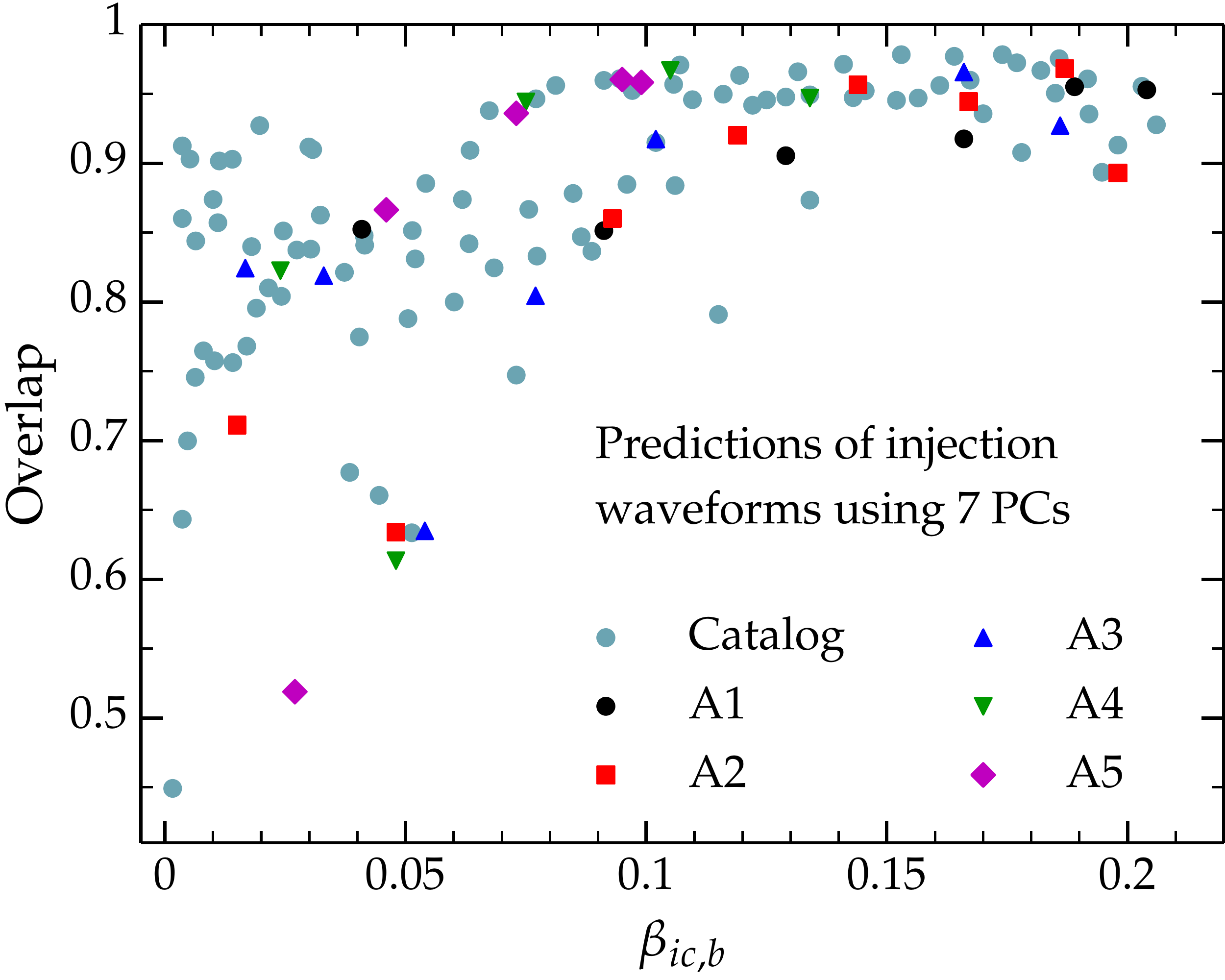}}
\caption{\label{fig:oos} \small Predictions of the 31
  Abdikamalov~\emph{et al.} \emph{injection} waveforms (see
  Sec.~\ref{sec:AbCat}) using the design matrix used to produce
  Fig.~\ref{fig:a7polyintera}.  For comparison, we include the catalog
  reconstructions from Fig.~\ref{fig:a7polyintera} marked as grey
  dots, denoted ``catalog'' in the legend.  We find that this
  particular model can predict injections waveforms very well, despite
  a few outliers.  }
\end{figure}
\begin{figure}[t!]
    \centerline{\includegraphics[width=8.6cm]{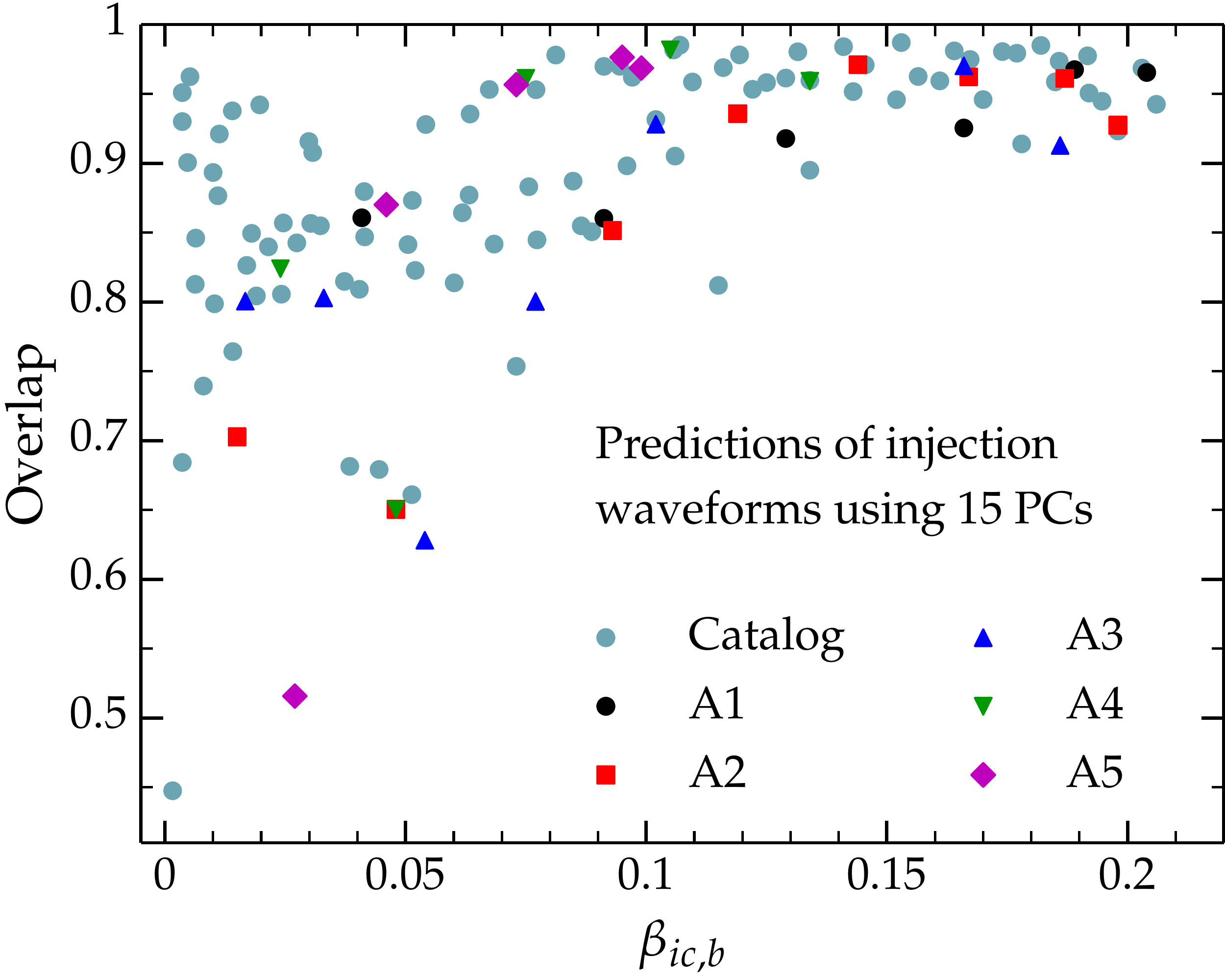}}
\caption{\label{fig:oos2} \small Predictions of the 31
  Abdikamalov~\emph{et al.} \emph{injection} waveforms (see
  Sec.~\ref{sec:AbCat}) using the design matrix used to produce
  Fig.~\ref{fig:a7polyintera}.  This plot was created identically to
  Fig.~\ref{fig:oos}, except 15 instead of 7 PCs were used to
  reconstruct the 91 catalog waveforms (gray dots) and predict the
  injection waveforms.  We find that using a larger number of PCs has
  little change on the reconstruction and prediction overlaps.  }
\end{figure}

To predict the subset of 31 injection waveforms, we employ our
previously fitted regression model whose design matrix was comprised
of a deviation encoding of $A$, a 5th order polynomial model on
$\beta_{ic,b}$, and two-way interactions between $A$ and
$\beta_{ic,b}$.  We use use Eq.~\ref{eq:predictioneq} to rapidly
generate these waveforms, given a vector, $\mathbf{\tilde{x}}$, of
their properly encoded physical parameters.

In Fig.~\ref{fig:oos}, we plot the overlap of the injections and their
predictions.  For comparison, we show in light grey dots
  the overlaps of the reconstructed waveforms of the original waveform
  set. These are copies of the markers shown in
  Fig.~\ref{fig:a7polyintera}.  The colored markers show the overlap
as a function of $\beta_{ic,b}$ of the 31 injection waveforms with
their predictions.  Many of the injection waveforms are predicted as
well as the waveforms in the original set are reconstructed.  The
presence of a few outliers (mostly at small to moderate $\beta_{ic,b}$)
indicates that there is room to improve our encodings of the physical
parameters.

Next, we reproduce Fig.~\ref{fig:oos} using 15 instead of 7 PCs in the
regression model.  Fig.~\ref{fig:oos2} shows that increasing the
number of PCs in our basis from 7 to 15 achieves only a
marginal increase in overlap for both the original and the injection
waveform sets.  This indicates that the first several PCs capture the
large majority of physically significant waveform content.  While
there is currently no clear rule that could guide us in
  choosing the appropriate number $k$ of PCs to use, we find that in this
context the choice of $k$ (as long as it is ``large enough'') has a
small impact on results.

\begin{figure}[t!] 
    \centerline{\includegraphics[width=8.6cm]{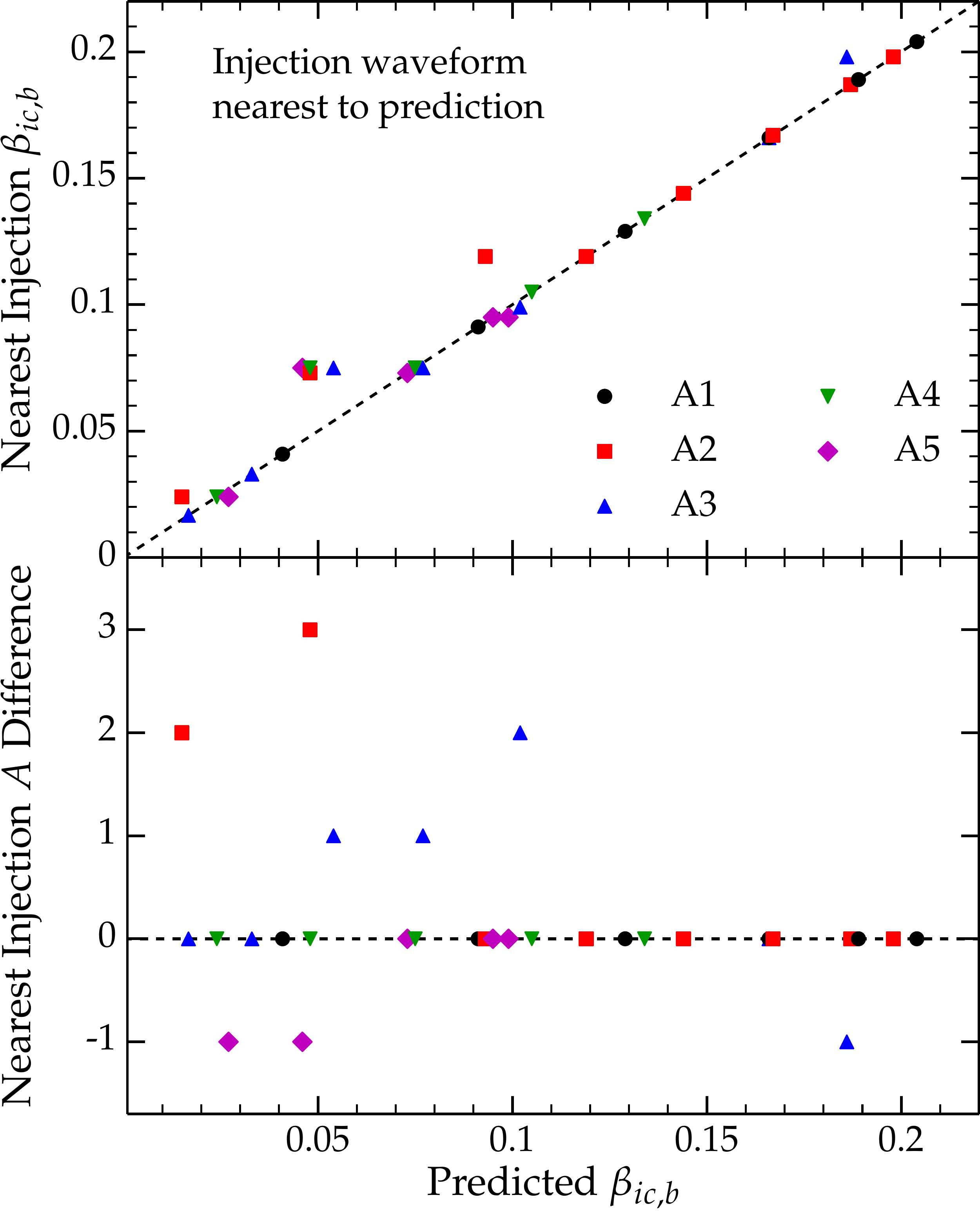}}
\caption{\label{fig:nearest} \small After predicting the 31 waveforms
  in the injection set, we mark the injection
    waveform that has the highest overlap with the predicted 
    waveform.  If the $i$th mark lies on the dotted black line, then
  the prediction of the $i$th injection waveform has the highest
  overlap with the $i$th injection waveform.   In the top
    panel, we plot the $\beta_{ic,b}$ of the nearest injection
    waveform versus the $\beta_{ic,b}$ value of the predicted
    waveform.  In the bottom panel, we plot the difference in $A$
    between the predicted waveform and the nearest injection 
    waveform as a function of $\beta_{ic,b}$.  } 
\end{figure}

We also test if the predicted waveform for the parameters associated
with a given injection waveform actually has its greatest overlap with
that waveform and not with some other waveform of the injection set.
In the top panel of Fig.~\ref{fig:nearest}, we mark the actual
injection waveform nearest to its prediction. We do this as a function
of the dominant parameter $\beta_{ic,b}$. If an injection has the
highest overlap with its prediction, then it is marked on the diagonal
dashed line.  We find that most of these marks lie on, or close to,
the diagonal. Hence, in most cases the predicted waveform is
identified with the injection waveform whose parameters where used for
its prediction.

In the top panel of Fig.~\ref{fig:nearest}, at $\beta_{ic,b} \approx
0.05$, four of the predictions are considerably nearer to the
$\beta_{ic,b} \approx 0.07$ injection waveforms.  Otherwise, only two
other injections have sub-optimal predictions, the $A2$, $\beta_{ic,b}
= 0.093$ and the $A3$, $\beta_{ic,b} = 0.186$ injection waveforms.  We
also note from the top panel of Fig.~\ref{fig:nearest} that the
prediction for the $A5$, $\beta_{ic,b} = 0.027$ injection waveform is
very near the diagonal, despite the fact that it has the lowest
overlap with its reconstruction in Figs.~\ref{fig:oos}
and~\ref{fig:oos2}.  Thus, its overlap with other injection waveforms
must be even lower.

In the bottom panel of Fig.~\ref{fig:nearest}, we plot the
$\beta_{ic,b}$ of the predicted injection waveform versus the
difference in $A$ between the predicted injection waveform and the
nearest injection waveform.  We note that for each instance where the
difference in $A$ is not equal to zero, the same waveform in the top
panel is marked off the diagonal.  Since there are only 31 injection
waveforms, a lack of overlap between the prediction and the injection
due to a problem fitting $\beta_{ic,b}$ results in $A$ being predicted
incorrectly, because $\beta_{ic,b}$ is the dominant parameter.  In
further work we plan on exploring different approaches to modeling the
waveforms' dependence on $\beta_{ic,b}$.

Figures~\ref{fig:oos},~\ref{fig:oos2}, and~\ref{fig:nearest} taken
together show that our regression approach produces good predictions
for $\beta_{ic,b} \gtrsim 0.06$ waveforms.  Potentially, waveform
dependence on rotation below $\beta_{ic,b} \approx 0.06$ is
inadequately fitted by a 5th order polynomial. In addition, the
appearance of postbounce prompt convection at slow to moderate rotation
and the associated appearance of stochastic GW signal features may
spoil our analysis.

\section{Summary and Further Work}

In this work, we have described a multivariate regression approach for
the analysis of simulated gravitational waveforms from rotating core
collapse.  The solutions of our regression model are \emph{feature
  vectors} --- pieces of waveform morphology \emph{directly}
attributable to encoded physical parameters.  While specific values of
discrete physical parameters are encoded individually, we have also
considered continuous parameter encodings to describe linear and
non-linear waveform dependence.

By constructing feature vectors from linear combinations of principal
components (PCs), we provided a means to connect the PC based methods
of previous work~\cite{Heng2009,Rover2009,Logue2012} to the physical
parameters underlying rotating core collapse.  Within the regression
framework, we use statistical hypothesis testing to quantitatively
measure how strongly feature vectors (thus physical parameters)
influence waveform morphology in the presence of Gaussian
  noise of a single gravitational-wave detector.

Finally, we used our regression model to \emph{reconstruct} and
\emph{predict} GWs from a given PC basis and set of encoded physical
progenitor parameters.  These reconstructions and predictions are
linear combinations of feature vectors, providing readily
interpretable solutions.  Our proof-of-principle study showed that our
regression scheme reliably interpolates between waveforms from
progenitors that have $\beta_{ic,b} \gtrsim 0.06$ (where
$\beta_{ic,b}$ is the ratio of rotational kinetic energy to
gravitational energy of the inner core at bounce).

We demonstrated our methodology on the recent Abdikamalov~\emph{et
  al.}~\cite{Abdikamalov2013} rotating core collapse waveform catalog.
Their core-collapse models are determined by two rotation parameters,
differential rotation ($A$) and $\beta_{ic,b}$.  Our statistical
hypothesis test based study of waveform parameter dependence
corroborates the more qualitative analysis
within~\cite{Abdikamalov2013}.  The axisymmetric simulations of
Abdikamalov~\emph{et al.}~\citep{Abdikamalov2013} produced linearly
polarized gravitational waveforms.  As full 3D models of
  stellar collapse and postbounce supernova evolution mature, we will
  need to adapt our regression scheme to handle waveforms with
  multiple polarizations and consider noise in gravitational-wave
  detector networks.

While we have shown that our regression strategy is effective for
rotating core collapse waveforms, it remains to test its ability
 on other gravitational-wave emission processes in
  stellar collapse and core-collapse supernovae.  For example, in the
  context of neutrino-driven explosions in nonrotating or slowly
  rotating progenitors, convective motions introduce stochastic
  components into the produced gravitational waves.  While able to
extract deterministic waveform features, our current regression model
cannot handle stochastic waveform components or varying degrees
of stochasticity dependent on progenitor parameters.

The primary focus of this work was on analyzing the relationships
between physical parameters and generated waveforms.  In the future,
we intend to shift our focus to waveform prediction in
  the context of parameter estimation for observed signals.  With the
rich statistical literature on regression modeling, there are many
avenues to explore.  We found that our waveform predictions using 5th
order polynomials of $\beta_{ic,b}$ are not as accurate
  for slowly and moderately rapidly rotating stellar cores with
  $\beta_{ic,b} \lesssim 0.06$.  Possibly, the degree of
stochasticity increases within cores at lower values of
$\beta_{ic,b}$.  Also, polynomials may not be the most effective basis
for expressing waveforms' dependence on $\beta_{ic,b}$.  Other bases,
such as splines or radial basis functions~\cite{Tibshirani} may
provide better fits.  Additionally, Gaussian Process regression
methods~\cite{GPforML} do not require one to specify a specific basis
for continuous physical parameters, and have been shown to capably fit
trends of arbitrary complexity.

 Multi-dimensional stellar collapse and core-collapse
  supernova simulations are still computationally challenging and time
  consuming. This currently prohibits the construction of dense
waveform catalogs exploring the full range of the physical parameter
space.  The ability to confidently predict waveforms
given an arbitrary set of parameter values (and a set of physical
parameters and waveforms that can be spanned by a PC basis) enables
template-bank based parameter estimation methods for linearly
polarized gravitational waves from rotating core
collapse.  In future work, this capability must be
  extended to include other important emission mechanisms, such as
  neutrino-driven convection, asymmetric neutrino emission, and
  nonaxisymmetric rotational instabilities.

\acknowledgements

We acknowledge helpful discussions with and help from members of the
LIGO Scientific Collaboration and Virgo Collaboration Supernova
Working Group, in particular Sarah Gossan, I. Siong Heng, and Nelson
Christensen.  BE and RF are supported in part by NSF grant
PHY-1205952.  CDO is partially supported by NSF CAREER grant
PHY-1151197, NSF gravitational physics grant PHY-0904015, The Sherman
Fairchild Foundation, and the Alfred P. Sloan Foundation. Some of the
computation performed towards the results presented here used NSF
XSEDE computing resources under award TG-PHY100033.


\end{document}